\documentclass[preprint,12pt]{elsarticle}
\usepackage{setspace}
\usepackage{amsthm}
\theoremstyle{definition}

\usepackage{graphicx}
\usepackage{dcolumn}
\usepackage{multirow}
\usepackage{subfigure}
\usepackage{times,mathptm}
\usepackage{float}
\usepackage{color}
\usepackage{amsmath,amsfonts}
\usepackage{mathptmx}
\usepackage{mathrsfs}
\usepackage{bbm}
\usepackage{bm}
\usepackage{xfrac}

\newcommand{\beq}{\begin{equation}}
\newcommand{\eeq}{\end{equation}} 
\newcommand{\bea}{\begin{eqnarray}}
\newcommand{\eea}{\end{eqnarray}}

\newcommand{\E}{\mathcal{E}}

\renewcommand{\d}{\delta}

\renewcommand{\ni}{\noindent}

\renewcommand{\o}{\omega}
\newcommand{\bx}{\mathbf{x}}

\newcommand{\bw}{\mathbf{w}}

\newcommand{\vx}{{\vec{x}}}

\newcommand{\m}{\mu}

\newcommand{\e}{\epsilon}
\newcommand{\s}{\sigma}

\newcommand{\D}{\Delta}

\newcommand{\oh}{\frac{1}{2}}

\newcommand{\dg}{\dagger}
\newcommand{\non}{\nonumber}

\newcommand{\rf}[1]{(\ref{#1})}
\newcommand{\ra}{\rightarrow}
\newcommand{\up}{\uparrow}
\newcommand{\dn}{\downarrow}

\renewcommand{\vec}[1]{\bm #1}

\usepackage{ulem}

\journal{Annals of Physics}

\begin{document}

\begin{frontmatter}

\title{Multiplicity, localization, and domains in the Hartree-Fock ground state of the two-dimensional Hubbard model} 

\bigskip
\bigskip

\author[1]{Kazue Matsuyama} 
\ead{kazuem@sfsu.edu}
\author[1]{Jeff Greensite}
\ead{greensit@sfsu.edu}
\address[1]{Physics and Astronomy Department, San Francisco State University, \\ 1600 Holloway Ave,
San Francisco, CA 94132 USA}
\begin{abstract}
 
   We explore certain properties of the Hartree-Fock approximation to the ground state of
the two-dimensional Hubbard model, emphasizing the fact that in the Hartree approach there is an enormous multiplicity of self-consistent solutions which are nearly degenerate in energy, reminiscent of a spin glass, but which may differ substantially in other bulk properties. 
It is argued that this multiplicity is physically relevant at low temperatures.  We study the localization properties of the one-particle wavefunctions comprising the Hartree-Fock states, and find that these are unlocalized at small and moderate values of $U/t$, in particular in the stripe region, but become highly localized at values corresponding to strong repulsion.  We also find rectangular domains as well as stripes in the stripe region of the phase diagram, and study pair correlations in the neighborhood of half-filling.
 
\end{abstract}

\end{frontmatter}

\bibliographystyle{elsarticle-num} 

\section{\label{Intro} Introduction}

    It is well known that the wave function of a particle in a stochastic potential may be localized; this is called
Anderson localization.  One of the questions we would like to address here is whether, in a
many-body quantum system with a translation-invariant Hamiltonian, such a phenomenon could also occur at zero or
very low temperature.  The rough idea is that the potential seen by one particle, due to all particles in the system,
appears to be disordered, perhaps sufficiently so as to induce localization.  It seems obvious that if a translation invariant
Hamiltonian in a finite volume has a unique ground state, then spatial localization in that state is impossible; the ground state would have to share the translation invariance of the Hamiltonian, and the expectation value of any observable must likewise be translation invariant.  But a different, spin-glass scenario is possible, in which the low temperature behavior is not
necessarily dominated by the lowest energy  state.  It may be that the landscape consists of
very many states which, as in a spin glass, are nearly degenerate in energy.  As a many-body system is cooled it may become stuck in one of these states, which need not be translation invariant, and in this scenario some analog of Anderson localization
could be possible.  This would answer in the affirmative the question raised in ref.\ \cite{Nandi}, of whether localization could occur in a
system described by a translation-invariant many-body Hamiltonian, with a stochastic element introduced only in the initialization.

    The picture of one particle moving in a random potential due to all other particles of the system immediately suggests
a Hartree-Fock approach, and the two-dimensional Hubbard model, which has been extensively studied by the Hartree-Fock
method, see e.g.\ \cite{Penn,Hirsch,Poilblanc,Zaanen,Machida,Schulz1,Schulz2,Ichimura,Verges,Inui,Dasgupta,Bach,Imada,Fazekas,Xu,multiref,PRX} 
seems like a good framework in which to address the questions we are interested in.  It is known \cite{Verges,Inui,Xu} that there is no unique solution of the self-consistent Hartree-Fock equations; rather there is great multiplicity of self-consistent solutions generated by an iterative
procedure.  Some authors (e.g.\ \cite{Xu}) look for states of lower energy out this multiplicity of states via some simulated annealing procedure.
We believe this is not the right approach if the system at very low temperature exhibits
localization or translation-symmetry breaking of any sort, and a multiplicity of states is physically relevant. The situation is then better described by analogy to spin glasses, where the
search for the global minimum of the energy seems futile, and we accept that the system, cooled from some finite temperature, will get stuck in one or another local minimum.  It is possible that none of the numerous self-consistent solutions in the Hartree-Fock approach represent the true, translation-invariant ground state, but rather correspond to very long-lived metastable states in the exact theory.  The example of the double well potential with a very high barrier between the two minima comes to mind.  The true ground state has the symmetry of the Hamiltonian, but there exist states which remain concentrated in one of the two wells for a long period which depends on the tunneling amplitude. In case of 2D many body systems with a translation invariant Hamiltonian, if spatial dependence in some observables is seen experimentally at very low temperatures, then it must be that a multiplicity of either stable, or long-lived metastable states, such as the states identified by the Hartree-Fock procedure, are the states which are physically relevant.

   Localization of one particle wavefunctions in the Slater determinant has not, to our knowledge, been studied in the context of the Hartree-Fock approximation to the standard Hubbard model (i.e.\ the model with no additional random potential).  In fact the standard 2D Hubbard model is known to exhibit periodic patterns such as stripes and a checkerboard in certain
regions of the phase diagram, and it is natural to attribute those patterns to localized spin up/down electrons at each site.  As we will see,
this is not the case; electron localization in the Hubbard model, as quantified by the inverse participation ratio (IPR), appears only at
rather large values of $U/t$.

   Apart from localization there are of course many other observables
which can be computed in the framework of the Hartree-Fock approach, and in this article we will display a selection which we feel are either relevant to the multiplicity of solutions, or which may not have been emphasized in previous studies.  In particular we point out
the existence of rectangular domains in spin density distributions, in addition to the stripe patterns noted long ago. 

   We motivate the Hartree-Fock approximation to the Hubbard Hamiltonian in a slightly unconventional manner, drawn from our picture of a single electron moving in a stochastic potential generated by all other electrons.  Let us suppose that we have $M$ electrons on a two dimensional ${L\times L}$ square lattice with periodic boundary conditions, interacting according to the usual Hubbard model Hamiltonian
\bea
H &=&  -t  \sum_{<xy>} \sum_{s=\up,\dn}  c^\dg(x,s) c(y,s)  \non \\ 
    & &  + U \sum_{x} c^\dg(x,\up) c(x,\up) c^\dg(x,\dn) c(x,\dn) \ ,
\eea
with the first term a sum over nearest neighors, and define the Hartree-Fock state $|\Phi \rangle = \Phi |0\rangle$ where 
\beq
  \Phi =  \prod_{i=1}^M \sum_{\bx_i} \phi_i(\bx_i) c^\dg(\bx_i) \ ,
\label{Phi}
\eeq
and where the single-particle states $\{\phi_i\}$ are all orthogonal.
Our notation is that ${\bf x}=(x,s)$ where $x$ labels the lattice site, and ${s= \{\up, \dn\}}$ is the spin.  Now we introduce
an operator for an auxiliary (and fictitious) electron interacting with the other $M$ electrons, and define the reduced one-particle
Hamiltonian $H[\Phi]$ with matrix elements
\beq
   H_{\bw_1,\bw_2}[\Phi] = \langle 0|\Phi^\dg c(\bw_1) H c^\dg(\bw_2) \Phi |0\rangle_{conn} \ ,
\eeq
where the subscript $conn$ means that we keep only contributions to the expectation value which contain 
the anticommutators of $c(\bw_1)$ and $c^\dg(\bw_2)$ with creation/destruction operators in $H$.  This is our
candidate for the Hamiltonian describing the propagation of a single electron in the average field of all other
electrons.  The approximation to the Hubbard model ground state is arrived at by an iterative procedure.  
Given $\Phi^{(n)}$ at the $n$-th iteration, we solve numerically the eigenvalue problem 
\beq
H[\Phi^{(n)}] \phi_i(\vx) = \e_i \phi_i(\vx)
\eeq
for the $M$ lowest energy states, 
and insert those eigenstates $\{\phi_i,~i=1,2,...M\}$
into \rf{Phi} to obtain the next approximation $|\Phi^{n+1}\rangle$ to the ground state at the ${(n+1)}$-th iteration.  The procedure terminates when
a certain convergence criterion, described below, is satisfied.

\section{Procedure}

It is straightforward, given $\Phi$, to find the non-zero matrix elements: \\

\ni\underline{$w_1,w_2$ nearest neighbors}
\bea
    H[\Phi]_{\bw_1,\bw_2} = -t \d_{s_1,s_2}
\eea
and we note again that periodic boundary conditions are imposed. \\

\ni \underline{$w_1=w_2=w$ same site}

\bea
    H[\Phi]_{w\up,w\up} &=& U\rho(w,\dn,\dn) \non \\
    H[\Phi]_{w\dn,w\dn} &=& U \rho(w,\up,\up)  \non \\
    H[\Phi]_{w\up,w\dn} &=& -U \rho(w,\dn,\up)    \non \\
    H[\Phi]_{w\dn,w\up} &=&- U \rho(w,\up,\dn) \ .
\eea
where
\bea
 \rho(w,\dn,\dn) &=&  \sum_{i=1}^M \phi_i^*(w,\dn) \phi_i(w,\dn) \non \\
 \rho(w,\up,\up) &=&  \sum_{i=1}^M \phi_i^*(w,\up) \phi_i(w,\up) \non \\
 \rho(w,\dn,\up) &=&  \sum_{i=1}^M \phi_i^*(w,\dn) \phi_i(w,\up) \non \\
 \rho(w,\up,\dn) &=&  \sum_{i=1}^M \phi_i^*(w,\up) \phi_i(w,\dn) \ .
\label{rho}
 \eea
The problem is then to find the eigenvalues and corresponding eigenvectors of a $2L^2 \times 2L^2$
sparse matrix, which can be handled by standard numerical software.\footnote{We use the Matlab {\bf eigs} function.}  It is important to point out here that the iterative procedure does not converge to a unique ground state, but depends
instead on the (random) initialization.  Energy densities are only very weakly dependent on the initialization; magnetization and spatial distributions have a far stronger dependence; we will return to this point below.  We initialize the system with a small, random choice of $\rho$, i.e.\ at each site $w$ we generate three uniformly distributed random numbers $r_1,r_2,r_3$ in the range $[0,1]$, and let
\bea
\rho(w,\up,\up) &=& a r_1 \non \\
\rho(w,\dn,\dn) &=& a r_2  \non \\ 
\rho(w,\up,\dn) &=& \rho(w,\dn,\up) = b(r_3-0.5) \ ,
\label{ranstart}
\eea
with $a,b$ taken to be small constants, e.g.\ $a=0.01, b=0.001$.  The choice, and even the order of magnitude of these
constants is not critical.  However, while each set of iterations converges to a solution with very nearly degenerate energy densities, typically differing by fractional deviations of order $O(10^{-4})$, solutions obtained with different stochastic starting points  vary widely in the spatial distribution of spin up and
spin down electron densities.  If this Hartree-Fock result reflects a true property of the 2D Hubbard model, then the enormous
multiplicity of nearly degenerate ground states is reminiscent of a spin glass.

After the first iteration, one usually finds equal numbers of electrons of either spin
\beq
            \sum_w \rho(w,\up,\up) =  \sum_w \rho(w,\dn,\dn) = \oh M \ ,
\eeq
but any small deviation from this sum rule is corrected by adding or subtracting a small constant of order $1/L^2$ to
the electron densities, prior to the next iteration.  Thus we are selecting for self-consistent solutions with equal numbers of up and down
spins.  The convergence criterion is that after $n$ iterations the mean square deviations of electron spin densities satisfy
\bea
         {1\over L^2} \sum_w (\rho^{(n)}(w,\up,\up) - \rho^{(n-9)}(w,\up,\up))^2  &<& \d^2 \non \\
         {1\over L^2} \sum_w (\rho^{(n)}(w,\dn,\dn) - \rho^{(n-9)}(w,\dn,\dn))^2  &<& \d^2 \ ,
\eea
where the $\rho^{(n)}$ denote densities obtained after the $n$-th iteration.  We choose ${\d=0.001}$.  We have investigated, at some
points in the $U-$density plane, the effect of decreasing $\d$ by one or two orders of magnitude.  This will be discussed below, but
in brief the only effect of decreasing $\d$ by an order of magnitude is to increase the number of iterations required by convergence modestly, at $U/t=3$, or by a factor of 4 or 5 at $U/t=30$.  But strengthening the convergence criterion in this way makes very little difference quantitatively, e.g.\ on the energy density, and no difference whatever qualitatively.

   In all our numerical computations we have taken $t=1$, so $U/t=U$ in the results reported below.  

\subsection{Relation to the mean field decomposition}

    With the subtraction of an (irrelevant) constant, the operator
\beq
     H_{mf} =  \sum_{\bw_1,\bw_2} c^\dg(\bw_1) H[\Phi]_{\bw_1,\bw_2} c(\bw_2)  - E_0
\eeq
is the standard mean field approximation to the Hubbard model Hamiltonian, where   
\beq
E_0 = U \sum_x  \Bigl(\rho(x,\up,\up)\rho(x,\dn,\dn) - \rho(x,\up,\dn)\rho(x,\dn,\up)\Bigr) \ ,
\eeq 
cf.\ Fazekas \cite{Fazekas}, and Lechermann in \cite{Pavarini:17645}.  Some authors, however, adopt a simpler 
mean-field decomposition.  Writing
\beq
n(x,\up) n(x,\dn) = c^\dg(x,\up) c(x,\up) c^\dg(x,\dn) c(x,\dn) \ ,
\eeq
where
\beq
          n(x,s) =  c^\dg(x,s) c(x,s)  \ ,
\eeq
the simpler decomposition is
\bea
& & \lefteqn{c^\dg(x,\up) c(x,\up) c^\dg(x,\dn) c(y,\dn)} \non \\
& & \qquad \ra n(x,\up) \langle n(x,\dn) \rangle + \langle n(x,\up)\rangle n(x,\dn) 
  -  \langle n(x,\up)\rangle \langle n(x,\dn) \rangle  \ , \non \\ 
\eea
and in our notation this amounts to dropping the spin-fllip terms  $H[\Phi]_{x\up,x\dn}$ and  $H[\Phi]_{x\dn,x\up}$ 
in $H_{mf}$.  This is what is done in the Hartree-Fock calculations of ref.\ \cite{Xu}, whose approach is in other respects quite 
similar to our own. Dropping the spin-flip terms is, however, a severe truncation of the standard mean field expansion, and is analogous
to dropping the exchange term in atomic physics

   We also make no attempt to select the ``best'' self-consistent solution via some annealing procedure, which aims to select solutions which are slightly lower in energy than solutions obtained without this procedure.  This reflects our opinion, already discussed above, that a multiplicity of physically relevant states, analogous to a spin glass, is a requirement if translation non-invariance is seen at low temperatures,
and therefore a state which is very slightly lower in energy than most Hartree-Fock states (which are all nearly degenerate in energy) is of
no special significance.

\section{Observables}
 
    Once a self-consistent solution is obtained, we compute:
    
 \begin{itemize}
 
 \item The energy density
\beq
\E = {1\over L^2}\sum_{i=1}^{M}  \e_i \ .
\label{energy}
\eeq

\item The energy gap (at zero temperature) between the highest energy occupied and next unoccupied states
\beq
         \D \e = \e_{M+1} - \e_M \ .
 \label{gap}
 \eeq
 
 \item Charge and spin densities
 \bea
            C(x) &=& \rho(x,\up,\up) + \rho(x,\dn, \dn) \non \\
            D(x) &=& \rho(x,\up,\up) - \rho(x,\dn, \dn)  \ .
\label{CD}
\eea

 \item Local magnetization
\bea
            m &=& {1\over 2L^2}  \sum_x \sum_{\m=1}^2 D(x) D(x+\hat{\m}) \ .
\label{m}
\eea
To be clear, $m$ is {\it not} the global magnetization; it is simply the corellator of nearest neighbor spin densities.  Antiferromagnetic regions of the lattice will have negative $m$, ferromagnetic are positive $m$, but it is understood that non-zero $m$ does not necessarily imply
long-range order.

\item Long range order.  In principle any regular arrangement of spins in the 2D Hubbard model, at any non-zero temperature, is a violation of the Mermin-Wagner theorem. Nevertheless, such arrangements have been observed on finite lattices in quantum Monte Carlo simulations at half-filling \cite{Varney}; this must be attributed to the very low (zero) temperature and finite volume.  A useful observable to probe periodicity, at least on the scale of the finite lattice, is
\bea
S(k) &=& {1\over L^2}\sum_{x,y} D(x) D(y) e^{ik\cdot(x-y)}  \non \\ 
&=& {1\over L^2} \widetilde{D}(k) \widetilde{D}^*(k) \ ,
\label{Sk}
\eea
where $\widetilde{D}(k)$ is the Fourier transform of the spatial spin distribution $D(x)$.  We have long range order (with the caveat just
mentioned), when $S(k)$ is concentrated at only a few values of $k$, typically resulting in either a checkerboard, stripe, or rectangular domain pattern.
 
 \item Localization.  This is our original motivation.  We can judge whether individual
 energy eigenstates $\phi_i(x,s)$ are localized by computing the inverse participation ratio (IPR)
 \bea
       IPR_i = \sum_x \sum_s (\phi^*_i(x,s) \phi_i(x,s))^2 \ ,
 \eea
where $IPR_i=1$ means that the one electron state is localized to a single site and definite spin, while $IPR_i \sim 1/L^2$ indicates that the positional probability density is spread evenly over most of the lattice area.

\item Momentum distribution.  We compute momentum-space occupation numbers
\bea
n(k) &=& {1\over 2L^2} \sum_{x,y,s} e^{ik\cdot(x-y)} \langle \Phi| c^\dg(x,s) c(y,s) |\Phi \rangle \non \\
        &=& {1\over 2L^2} \sum_{i,s} \phi_i^*(k,s) \phi_i(k,s) \ ,
\eea
where
\beq
  \phi_i(k,s)  = \sum_x e^{ik\cdot x} \phi_i(x,s) \ ,
 \eeq
 and we also compute the magnitude of the discretized gradient
 \bea
 \nabla_x n(k) &=& {L \over 4 \pi} (n(k_x+1,k_y) - n(k_x-1,k_y) )\non \\
  \nabla_y n(k) &=& {L \over 4\pi} ( n(k_x,k_y+1) - n(k_x,k_y-1) )\non \\
 |\nabla n(k)| &=& \sqrt{(\nabla_x n)^2 + (\nabla_y n)^2} \ .
 \eea

\item Pairing correlations.  We search for d-wave arrangements in the correlator
\beq
          \Delta(k',k) = \langle c^\dg(k',\up) c^\dg(-k',\dn) c(k,\up) c(-k,\dn) \rangle \ ,
\label{pcorr}
\eeq
where $k'\ne k$, and choosing the $x,y$ components of $k$ to be the $y,x$ components of $k'$.   
 
 \end{itemize}

 \section{Results}

 \subsection{Momentum space distributions}

 A good starting point is to compare our results for momentum space distributions $n(k)$ and gradients  $|\nabla n(k)|$ with the results of Monte Carlo simulations, particularly where those Monte Carlo results exist away from half-filling.  Because of the sign problem, Monte Carlo simulations away from half-filling must rely on a reweighting procedure of some kind \cite{Blank}, and this apparently does not work down to zero temperature.  Thus our comparison with the Monte Carlo simulations of Varney et.\ al \cite{Varney}, away from half-filling, is necessarily a comparison of our zero temperature results with Monte Carlo simulations at finite temperature.  The comparison is nonetheless interesting, even if only at the qualitative level.  

  \begin{figure}[htb]
 \includegraphics[scale=0.8]{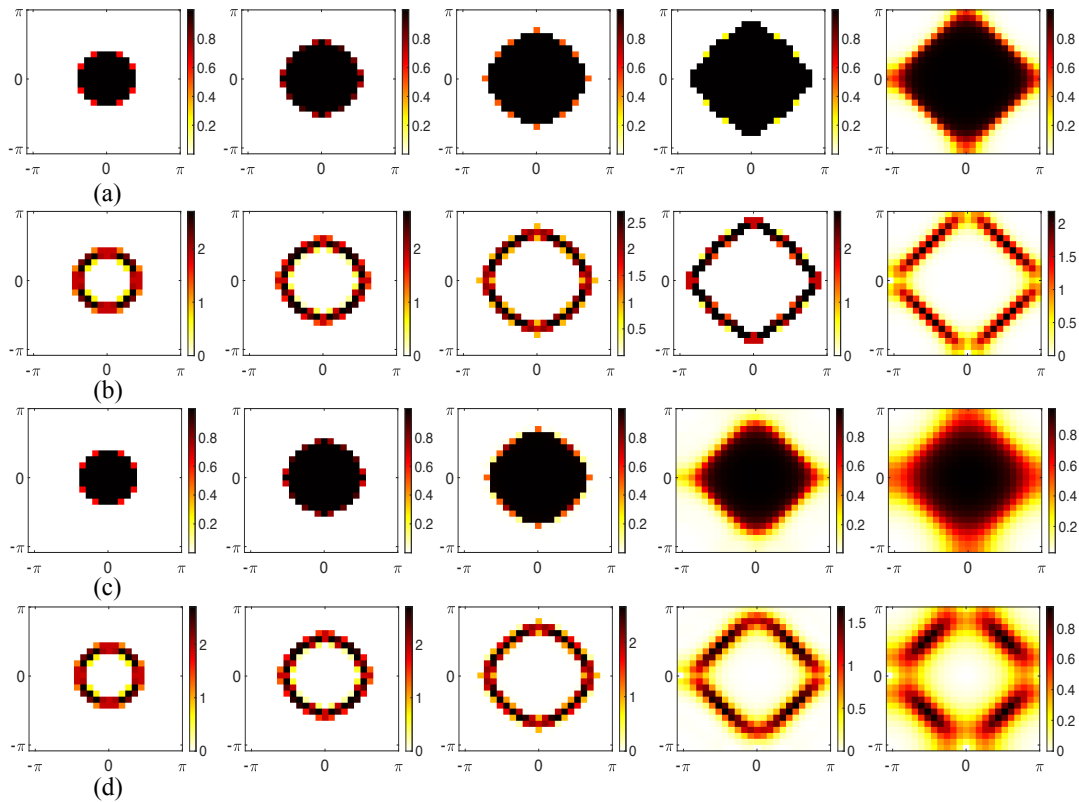}
 \caption{2D color plots of the momentum distribution $n(k_x,k_y)$ and the magnitude $|\nabla n|$ of its gradient. This plot should
be compared with the Monte Carlo results displayed in Figure 2 of Varney et al.\ \cite{Varney}.  Densities, increasing horizontally from left to right, are $f = 0.23, 0.41, 0.61, 0.79, 1.0$, where $f$ is the density.  (a) $n(k)$ at $U/t=2$; (b) $|\nabla n|$ at $U/t=2$;
(c) $n(k)$ at $U/t=4$; (d) $|\nabla n|$ at $U/t=4$. } 
\label{Varney}
 \end{figure}

    In Fig.\ \ref{Varney} we display our results (on a $24^2$ lattice) for momentum distributions $n(k)$ and gradients $|\nabla n(k)|$ at 
the same densities (electrons/site) \newline ${f=0.23, 0.41, 0.61, 0.79, 1.0}$  used in ref.\ \cite{Varney}, and Fig.\ \ref{Varney} should be compared
with Fig.\ 2 in that reference.  Despite the finite temperatures used in \cite{Varney} the figures are very similar, even at the quantitative level.  Note in particular the
absence of a sharp boundary in momentum space, separating occupied and unoccupied states at the larger filling
fractions, as $U$ is increased from $U=2$ to $U=4$.  
 
\subsection{Multiplicity of self-consistent states} 
 
\begin{figure}[t!]
\subfigure[~] 
{   
 \includegraphics[scale=0.5]{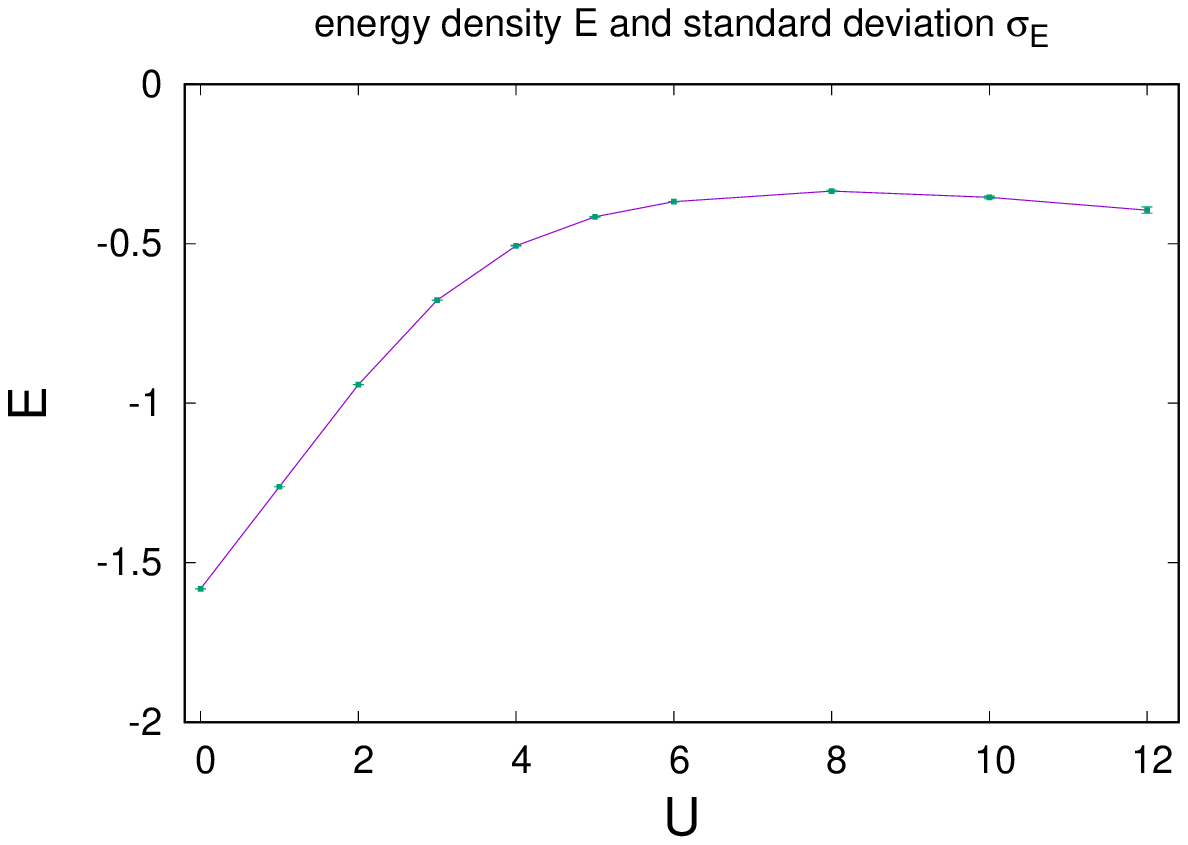}
 \label{std1}
}
\subfigure[~]
{   
 \includegraphics[scale=0.5]{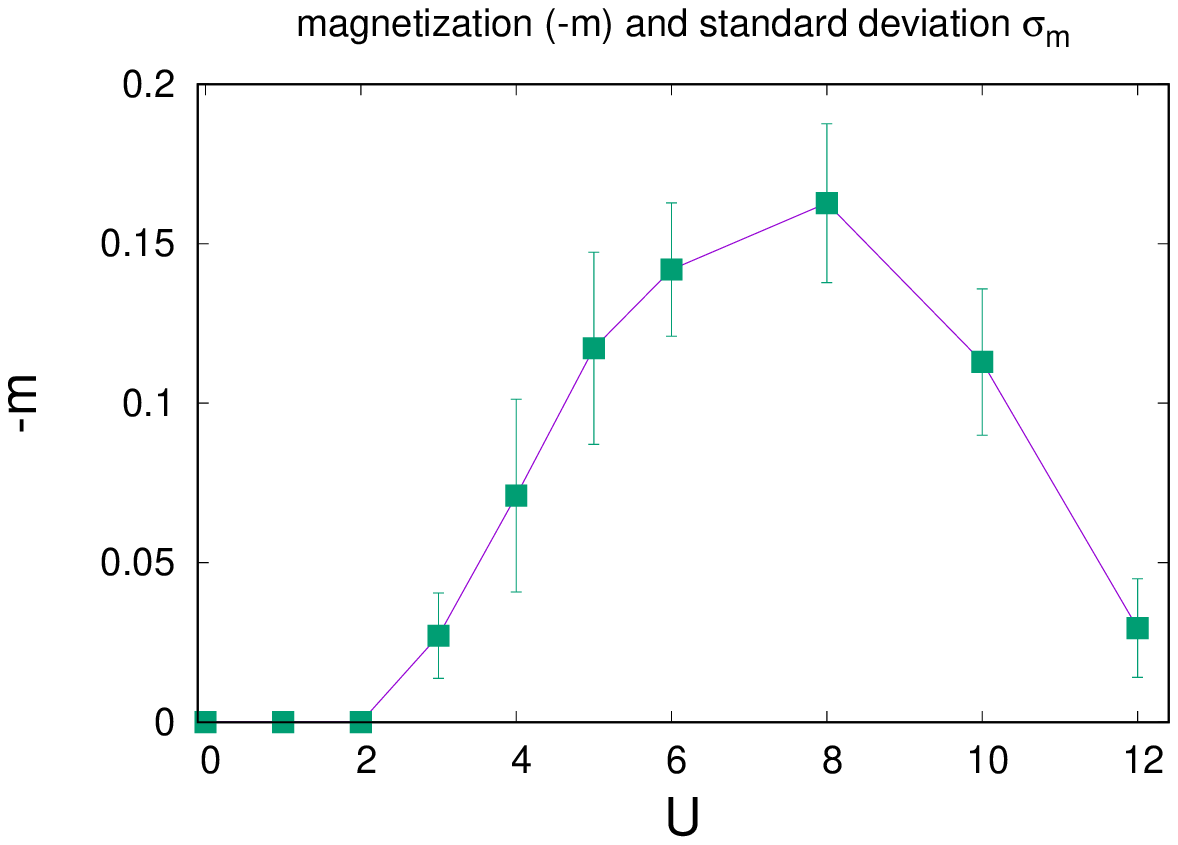}
 \label{std2}
}
\subfigure[~]
{   
\centerline{\includegraphics[scale=0.5]{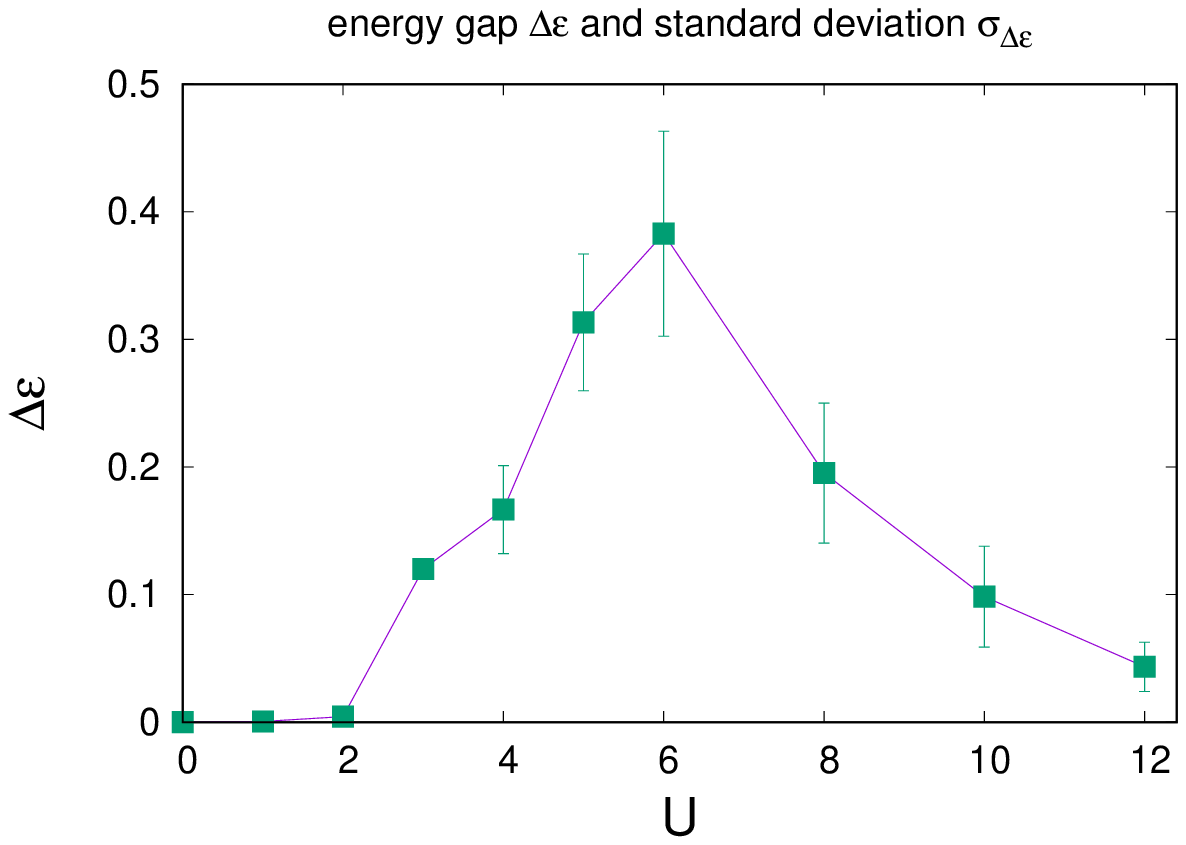}}
 \label{std3}
}
\caption{Values and standard deviations vs.\ $U$ at density $f=0.8$  of (a) energy density $\E$; (b) local magnetization $-m$; and (c) energy gap $\D \e$.  Standard deviations, displayed as error bars, were taken from a set of twenty independent, self-consistent Hartree-Fock solutions.  Note that no deviation is visible among these different solutions in the energy density, but there are significant deviations in
the magnetization and energy gap for $U\ge 3$.}
\label{stdev}
\end{figure}

   One significant aspect of the Hartree-Fock approach to the 2D Hubbard model is that there is no unique self-consistent solution for
the ground state, as noted many times in the literature.  Instead there are very many such states, obtained from different random starts \rf{ranstart}, except, of course, at $U=0$.  In Fig.\ \ref{stdev} we show our results at density $f=0.8$ on $24^2$ lattices for the energy density $\e$, the local magnetization $m$, and the energy gap $\D \e$, together
with their standard deviations.  The standard deviations are obtained at each $U$ from 20 separate self-consistent states, obtained
as described in section II.  These are plotted in Fig.\ \ref{stdev} as error bars, but we stress that in these figures the ``error bar''  represents the  standard deviation, rather than standard deviations of the mean.

   The standard deviation of energy density, $\s_\e$, is so small that it is not discernible in Fig.\ \ref{std1}.  As in a spin glass, with a landscape of local minima of near-degenerate energies, the different self-consistent solutions are nearly degenerate in energy.  The difference between these
self-consistent states is quite apparent, however, in the local magnetization $m$ and the energy gap $\D \e$, seen in Figs.\ \ref{std2} and \ref{std3} respectively.  In these cases, at $f=0.8$
and $U\ge 3$ (where local antiferromagnetic order becomes apparent), the standard deviations are quite substantial, indicating that
despite their near-degeneracy in energy, these different self-consistent solutions are physically distinct.
 
 \begin{figure}[htb]
\begin{center}
\subfigure[~]  
{   
 \includegraphics[scale=0.35]{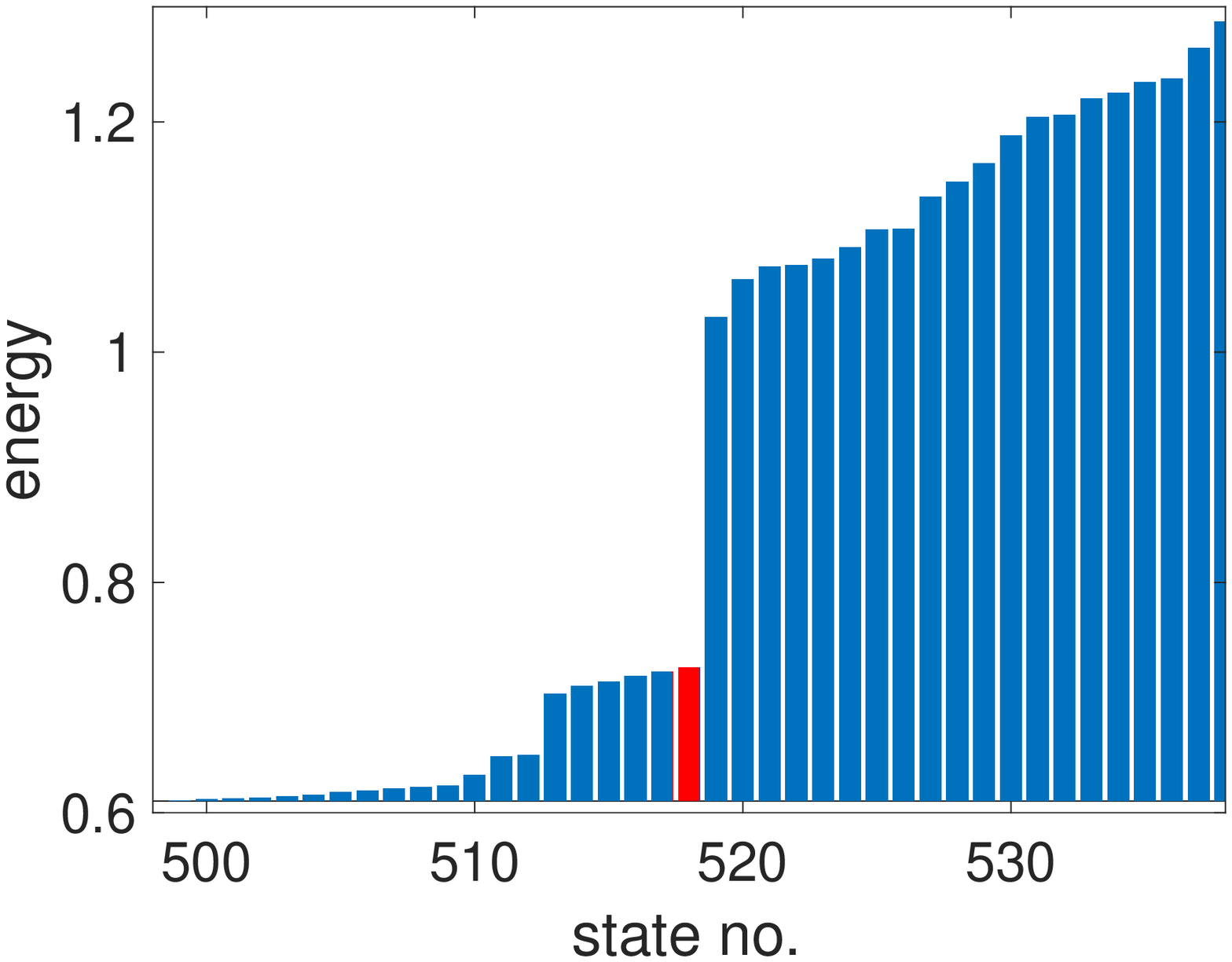}
 \label{gapA}
}
\hfill
\subfigure[~]
{   
 \includegraphics[scale=0.35]{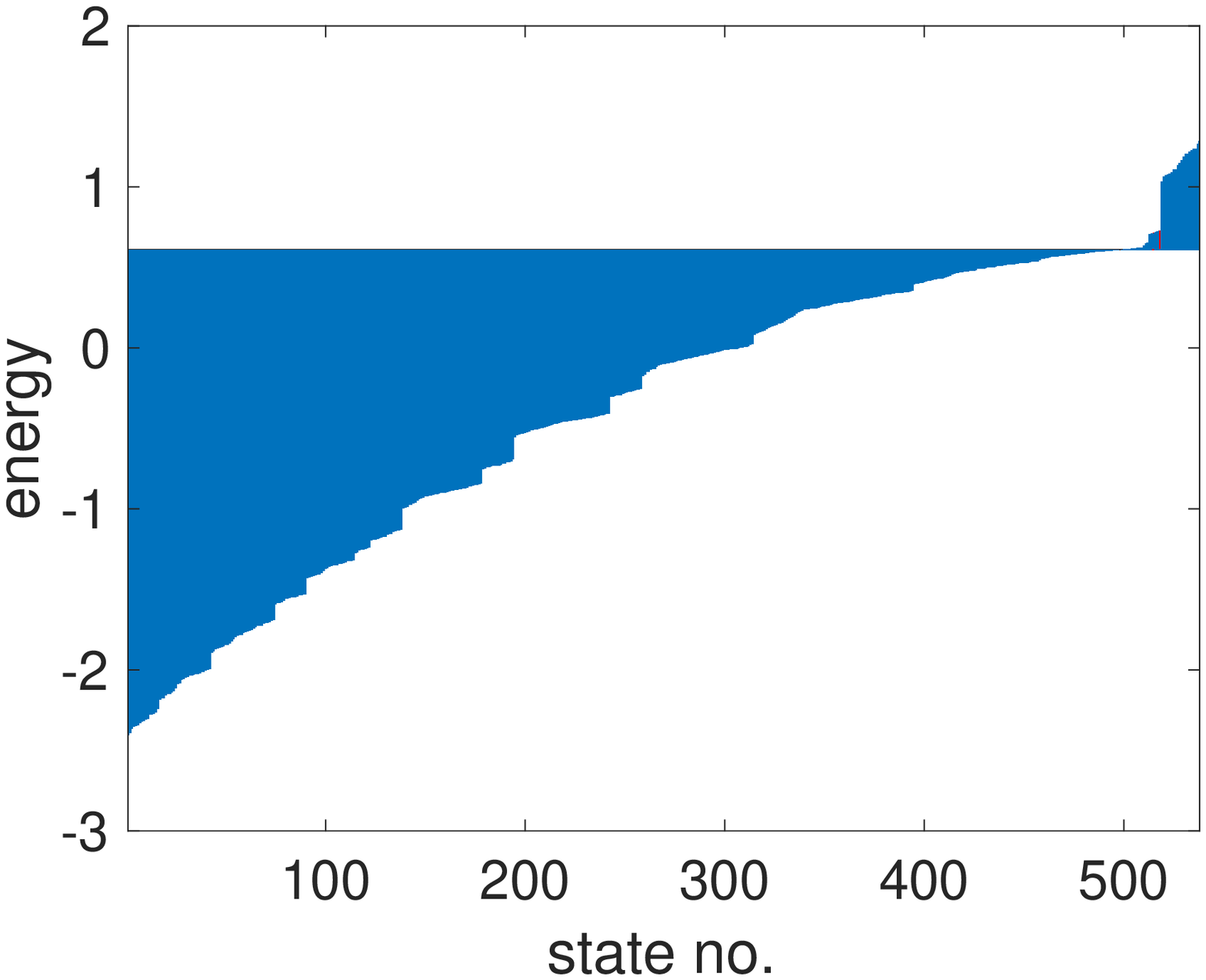}
 \label{gapB}
}
\end{center}
\caption{A histogram of energy vs.\ state number, with the bar for the last occupied state $\e_M$ shown in red. $U=4, f=0.9$ on a $24^2$ lattice.  (a) A closeup of energies in the region of the last occupied state; here the gap to the first unoccupied state is quite clear.  (b) Overview of energies $\e_i$ up to, and a little beyond, the last occupied state.}
\label{gaps}
\end{figure}
 \subsection{Local magnetization and the energy gap}
 
    In general there is a correlation, at zero temperature, between local magnetization $m$ and the gap $\D \e$ between the
energies of the last occupied and first unoccupied states.  These gaps increase with $|m|$, which in turn is largest at or near half-filling, and large $U$.  In Fig.\ \ref{gapA} we display a histogram of energies $\e_n$ vs.\ $n$ on a $24 \times 24$ lattice at $U=4$ and density $f=0.9$, taken from a typical configuration which has converged
from a random initialization, as described above.   This is a closeup view of energies in the neighborhood of $n=M$, with the energy of the last occupied state ($n=M=518$ for these parameters) shown in a different color.   Note the significant jump in energy between the last occupied state and the first unoccupied state ($n=M+1$). Figure \ref{gapB} is a more global display of the energies, beginning at the lowest energy.  

 \begin{figure}[htb]
 \centerline{\includegraphics[scale=0.7]{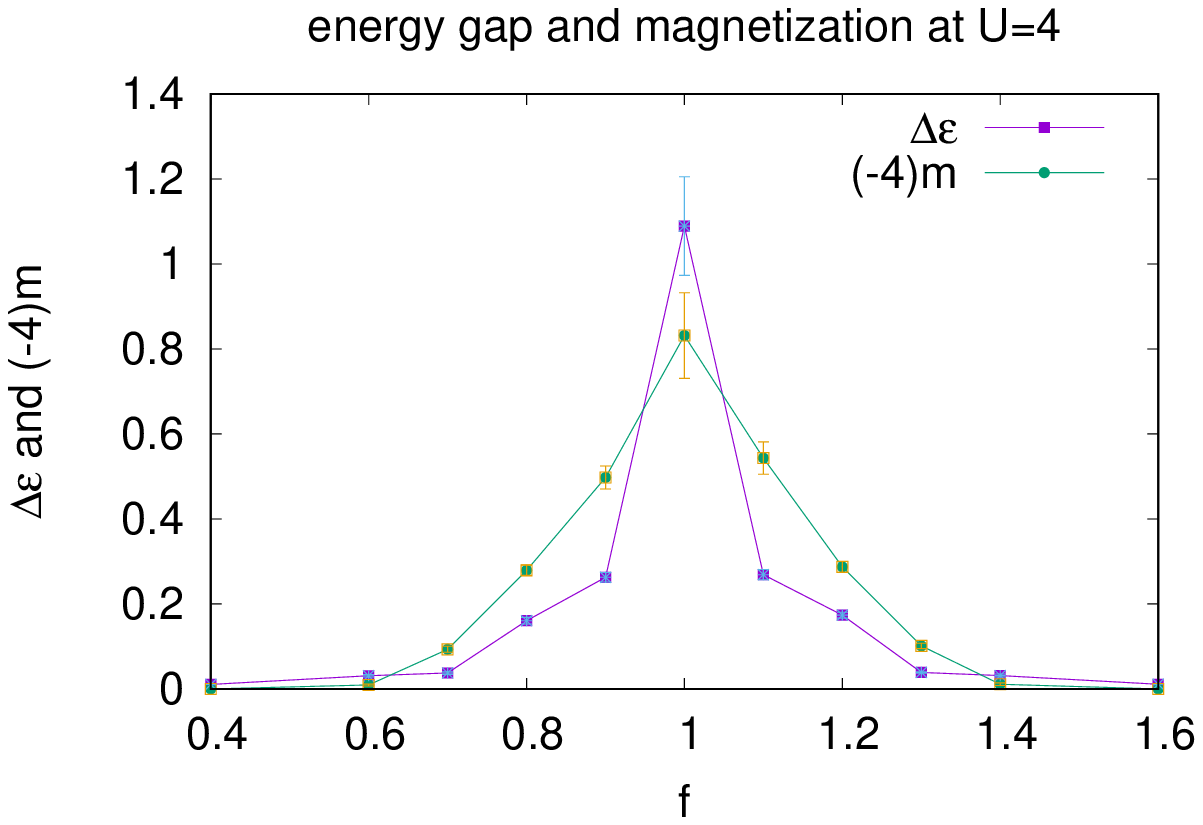}}
 \caption{Energy density and magnetization vs.\ density $f$ at ${U=4}$ on a $24^2$ lattice.  In this plot we have multiplied
 the magnetization by $-4$, for ease of comparison.}
 \label{gapmag}
 \end{figure}
\clearpage
\begin{figure}[htb]
\subfigure[~$(-1) \times$ local magnetization]  
{   
 \includegraphics[scale=0.33]{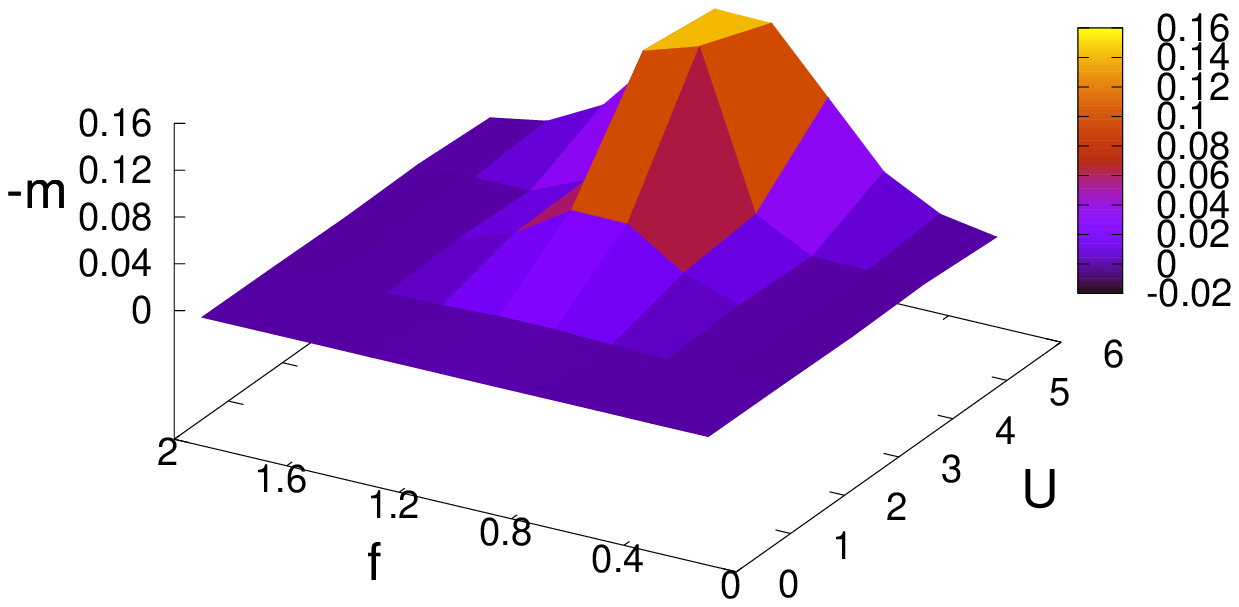}
 \label{mag3d}
}
\subfigure[~energy gap]
{   
 \includegraphics[scale=0.33]{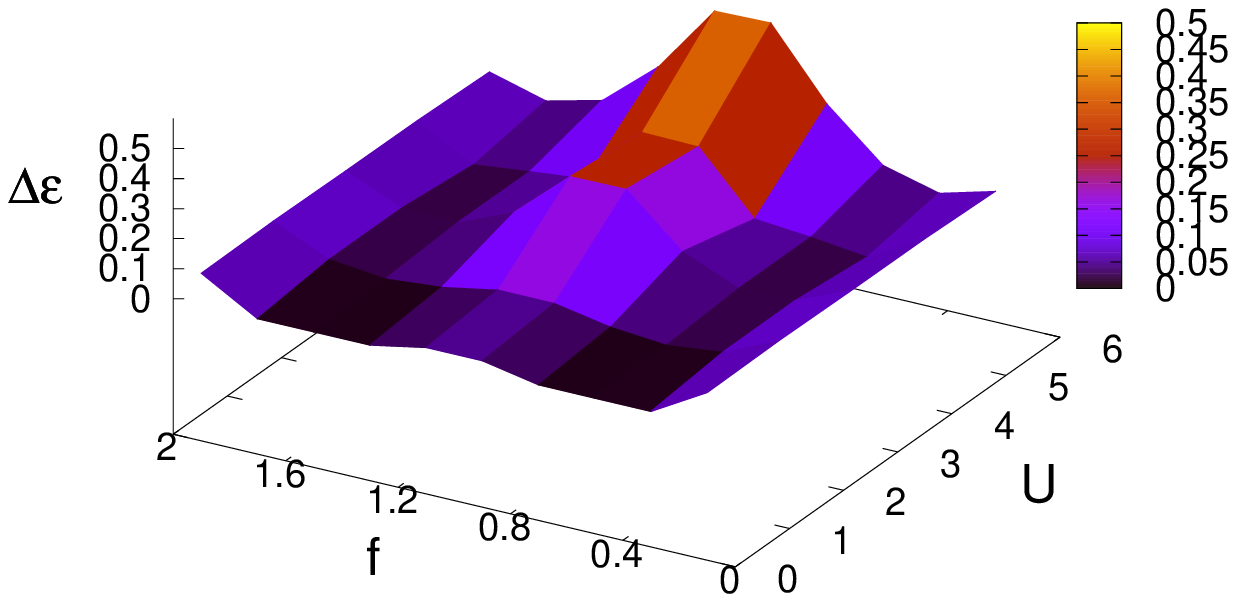}
 \label{gap3d}
}
\subfigure[~ energy density]
{   
 \includegraphics[scale=0.33]{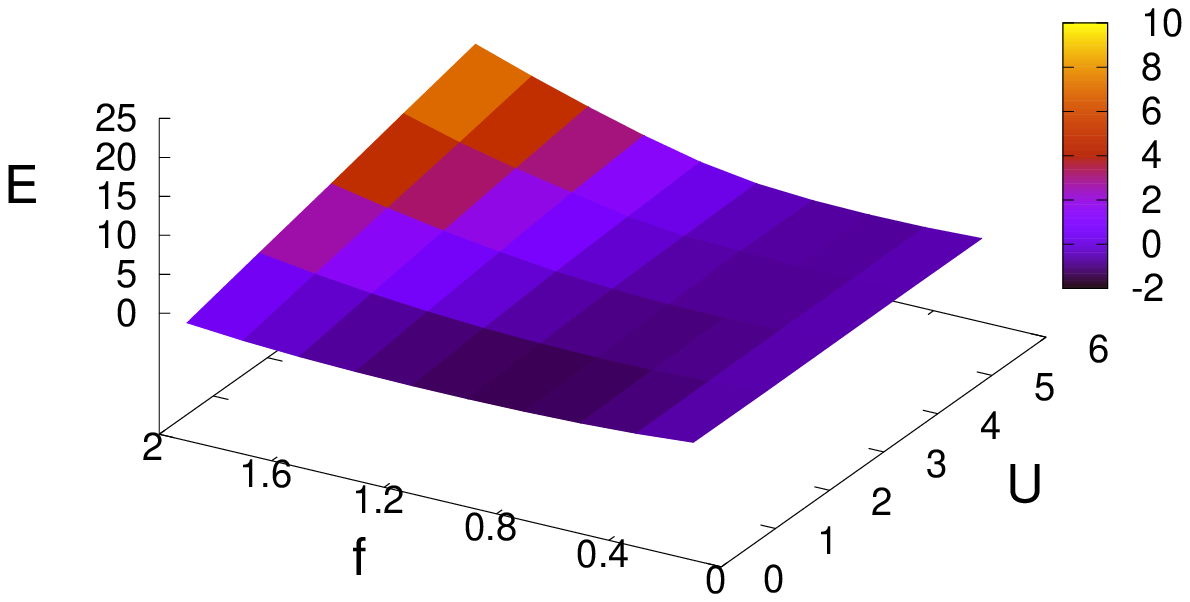}
 \label{E3d}
}
\caption{Surface plots of (a) local magnetization $\times (-1)$; (b) energy gap $\D\e$; (c) energy density $\E$; vs.\ coupling $U$ and density $f$.  All data taken on $30^2$ lattices. Note, in (a,b), that the region of significant local antiferromagnetic order is also the
region where the energy gap is significant.}
\label{view3d}
\end{figure}
    At moderate values of $U$, in a region centered around half-filling ($f=1$), we find that the local magnetization defined
in \rf{m} is significantly non-zero and negative, indicating local antiferromagnetism.  It is interesting that the local magnetization is closely correlated with the existence of an energy gap $\D \e > 0$.  An example of this correlation, at $U=4$ on a $24^2$ lattice, is shown in Fig.\ \ref{gapmag}.  In order that both the magnetization and energy gap are clearly visible on the same figure, we have
multiplied the magnetization by a factor of  $-4$.

\begin{figure}[h]
\centerline{\includegraphics[scale=0.4]{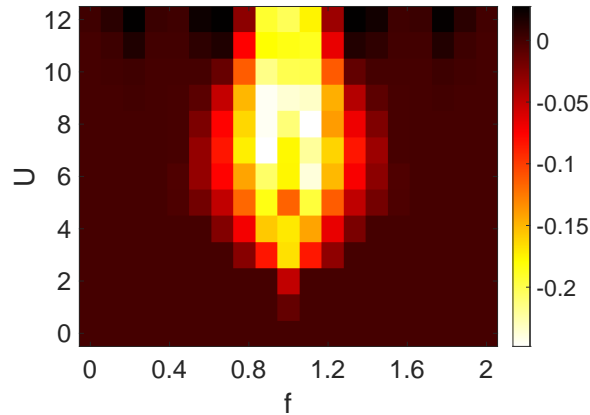}}
\caption{Color plot of local magnetization $m$ in the $f-U$ plane up to $U=12$. The bright region is a region of
local antiferromagnetism, surrounded by a region of negligible magnetization.}
\label{mg}
\end{figure}

\begin{figure}[htb]
\centerline{\includegraphics[scale=0.6]{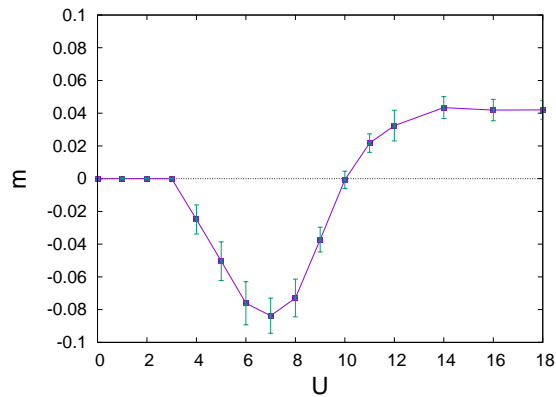}}
\caption{Local magnetization $m$ vs.\ $U$ at fixed filling density $f=0.7$.  We note that the local magnetization switches from
antiferromagnetic to ferromagnetic around $U=11$.  The ``error bars'' shown actually represent the standard deviation taken from
20 independent, self consistent Hartree-Fock solutions.}
\label{ferro}
\end{figure}      
      
      A more complete picture of the local magnetization $m$ and energy gap $\D \e$ in the $U-f$ parameter plane up to $U=6$ is
shown in Figs.\ \ref{mag3d} and \ref{gap3d}.  Again we see that local antiferromagnetic order in some region of the phase diagram is correlated with the existence of an energy gap $\D \e$.  The energy density in the $U-f$ plane is shown in Fig.\ \ref{E3d}.  There is
no obvious indication in this figure of any non-analyticity indicating a thermodynamic phase transition, although other types of
phase transitions (see, e.g., \cite{Kertesz}) may exist.  We note in passing that the energy density at $U=0$ is symmetric around
half-filling, as can be verified analytically.

     Fig.\ \ref{mg} is a 2D color plot of the local magnetization in the $f-U$ plane up to $U=12$.  Data is taken from the average of
results obtained from 20 independent Hartree-Fock ground state solutions on a $30^2$ lattice.  The brightly colored oval region is a
region of local antiferromagnetism; the local magnetization is negligible outside this region.  At the upper border of the diagram, at $U=12$, the local magnetization turns positive away from half-filling, e.g.\ at $f=0.7,1.3$ which marks the beginning of ferromagnetic regions in the phase diagram.  In Fig.\ \ref{ferro} we plot magnetization vs.\ $U$ at fixed $f=0.7$, also averaging results taken from 20 solutions on a $30^2$ lattice (again with ``error bars'' representing standard deviations) and it is clear that at this density, from $U=11$ onwards, the local magnetization is ferromagnetic rather than antiferromagnetic.  All in all, the picture presented in Figs.\ \ref{view3d} and \ref{mg} is  consistent, at least qualitatively, with much earlier Hartree-Fock explorations of the phase structure of the 2D Hubbard model, e.g.\ \cite{Hirsch}

\subsubsection{Large $U$ limit}

\begin{figure}[h]
\centerline{\includegraphics[scale=0.3]{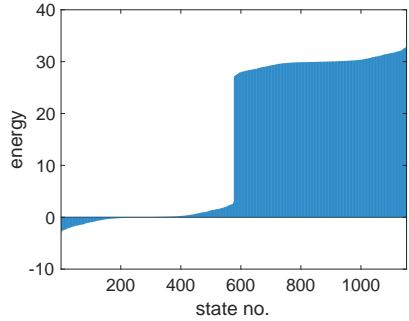}}
\caption{One particle energies at $U=30,~f=1$. The energy gap  between the last occupied and first unoccupied states is  $\D \e=24.6$.}
\label{gap30}
\end{figure}

\begin{figure}[h]
\centerline{\includegraphics[scale=0.5]{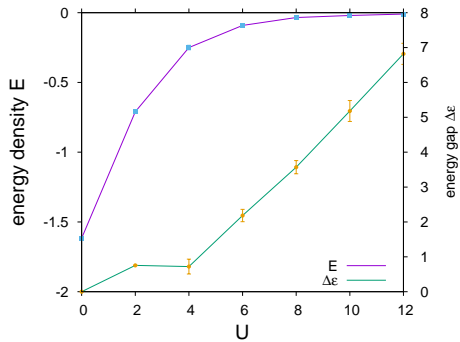}}
\caption{Convergence to $\E=0$, and linear growth of the energy gap $\D \e$ with increasing $U$, at
half-filling.}
\label{strong}
\end{figure}

   At $U=0$ the energies of one-particle states can be computed analytically,
\bea
          \e(m,n) &=& -2( \cos(2\pi n/L) + \cos(2\pi m/L) )  \non \\
          & & \text{where} ~~ m,n = 0,1,...,L-1 \ ,
\eea
and, after sorting these values from lowest to highest, agreement with the numerical calculation at $U=0$ is simply a modest check of
our code.   A more interesting limit is the computation of energies and
energy gap at half-filling and large $U$.  If we ignore the hopping term by comparison with the potential then the energy of the ground
state at the classical level is zero (each site occupied by a single electron), and the energy of the first excited state is simply $U$, corresponding to having one doubly-occupied site. The actual values for the $\e_i$ at $f=1$ and $U=30$ are shown in Fig.\ \ref{gap30}.  The last occupied state on this $24^2$ lattice
is at state number $M= 24^2 = 576$, and the energy gap in this case is found to be $\Delta \e = 24.6$, which is not so far from $U=30$.
The energy density of the ground state is very nearly zero, due to a near-exact cancellation of the positive and negative energies of the
occupied states.  The convergence to $\E=0$ at half-filling with increasing $U$, and the near linear increase of $\D\e$ with $U$ for
$U>4$ is seen in Fig.\ \ref{strong}.

     Subsequent figures will display the results (spin distribution, IPR values, etc.) taken from ``typical'' self-consistent solutions of the Hartree-Fock equations although, as seen above and also in the next section, ``typical'' solutions can vary a great deal in certain observables, spatial ordering in particular.
      
\subsection{Spatial patterns in the spin distribution}

\begin{figure}[t!]
 \centerline{ \includegraphics[scale=0.7]{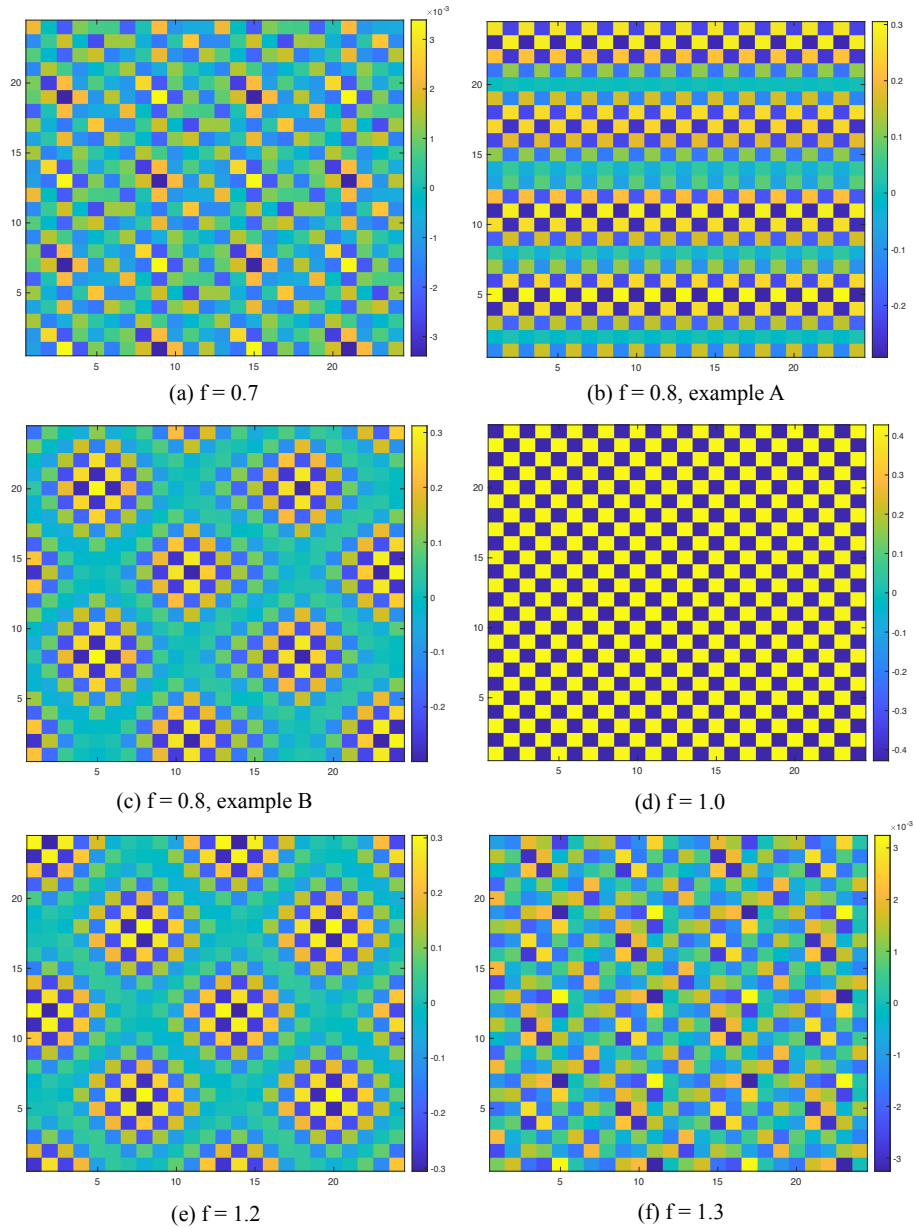}}
 \caption{Display of $D(x)$ at densities $f=0.7$ to $f=1.3$ and $U/t=3$.  The checkerboard pattern at $f=1$ (d) indicates an
antiferromagnetic order across the $24^2$ lattice area.  Within the antiferromagnetic region, e.g.\ at $f=0.8$ and $f=1.2$,
both ``wavelike'' in (b) and ``domains'' shown in (c) and (e) are found, depending on the initialization.  Outside the antiferromagnetic region there is no obvious pattern, and also the magnitude of $D(x)$ (note the colorbar scale at (a) and (f)) is reduced by two orders of magnitude.}
\label{geospin}
\end{figure}

   Examination of the spatial distribution of charge and spin densities, $C(x)$ and $D(x)$ respectively in \rf{CD}, reveal some
interesting geometric patterns in the neighborhood of half-filling.   The existence of stripes in Hartree-Fock treatments of the 2D Hubbard model goes back to \cite{Poilblanc,Zaanen,Machida,Schulz1}, and there is also experimental evidence of stripe order in strongly correlated materials \cite{Kivelson2,tranquadal}.
We will focus here on the spin densities $D(x)$, which are displayed in Fig.\ \ref{geospin} on a $24^2$ lattice at $U/t=3$ at zero temperature near half filling.  

\begin{figure}[htb]  
 \centerline{\includegraphics[scale=0.3]{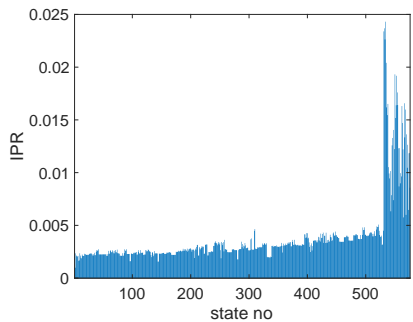}}
 \caption{Localization (IPR values) of one particle energy eigenstates $\phi_i$ vs.\ $i$ at $U/t=3, f=1$, $24^2$ lattice.  Despite the checkerboard pattern seen at these values in Fig.\ \ref{geospin}, the one particle wavefunctions are either not at all, or only very weakly localized.  Localization is a feature of strong repulsion (see Fig.\ \ref{IPR100} below), and is not associated with the stripe, domain, or checkerboard spatial patterns seen at moderate repulsion.}
 \label{ipr1U3}
\end{figure}

   At exactly half-filling one observes an antiferromagnetic ``checkerboard'' pattern, as seen in Fig.\  10(d). 
 Despite this apparently antiferromagnetic order, one should refrain from the interpretation that each electron is localized at one lattice site, in a pattern of alternating up/down spins.  In fact this is far from the case, as is best seen from an examination of the IPR of each $\phi_i$ at $U/t=3, ~ f=1$, which is shown in Fig.\ \ref{ipr1U3}.  The great majority of these IPR values are $\approx 0.0025$, and we recall that IPR=1 means that a particle is localized at a single lattice site, while IPR$=1/L^2$, which is 0.0017 on a $24^2$ lattice, is complete delocalization.  It is evident that almost all single electron states extend over the entire lattice, and the picture of electrons localized at sites in an up/down alternating pattern, as strongly suggested by the checkerboard pattern, is completely untenable.  It is remarkable that such highly
unlocalized electrons nonetheless contrive to produce such a regular geometric pattern.

\clearpage

   Stripe order in the spin density observable is seen in Fig.\ 10(b) 
at $f=0.8$.  This type of pattern is observed in
many of the self-consistent Hartree-Fock solutions generated by our iterative procedure with random initial conditions, at
moderate couplings in the neighborhood of half-filling.  But it is not the only pattern found.  An equally common pattern is the
periodic arrangement of quasi-rectangular domains seen in Fig.\ 10(c) 
again at $f=0.8$.  We emphasize that
these different patterns are obtained at the same coupling parameters and at the same zero temperature.  This is simply
another feature of the multiplicity of self-consistent Hartree-Fock solutions, as already discussed, and is again indicative of just how different these different solutions can be.   The possibility of domain structures of this kind was suggested previously in \cite{Fine}.

   We have examined, at $f=0.8$ and $U/t=3$, the sensitivity of our results to the convergence criterion.  Of course the number of  iterations
required to reach convergence varies significantly from one initialization to the next, but the following numbers are typical:  120 iterations to convergence on a $24^2$ lattice at $\d=10^{-3}$, 170 iterations at  $\d=10^{-4}$, 280 iterations at  $\d=10^{-4}$.  What is crucial is that the same geometric patterns, with the same energy density and very nearly the same amplitudes, are found in all three cases.  There is no
qualitative and very little quantitative difference.

\begin{figure}[t]
\begin{center}
\subfigure[~stripe]  
{   
 \includegraphics[scale=0.2]{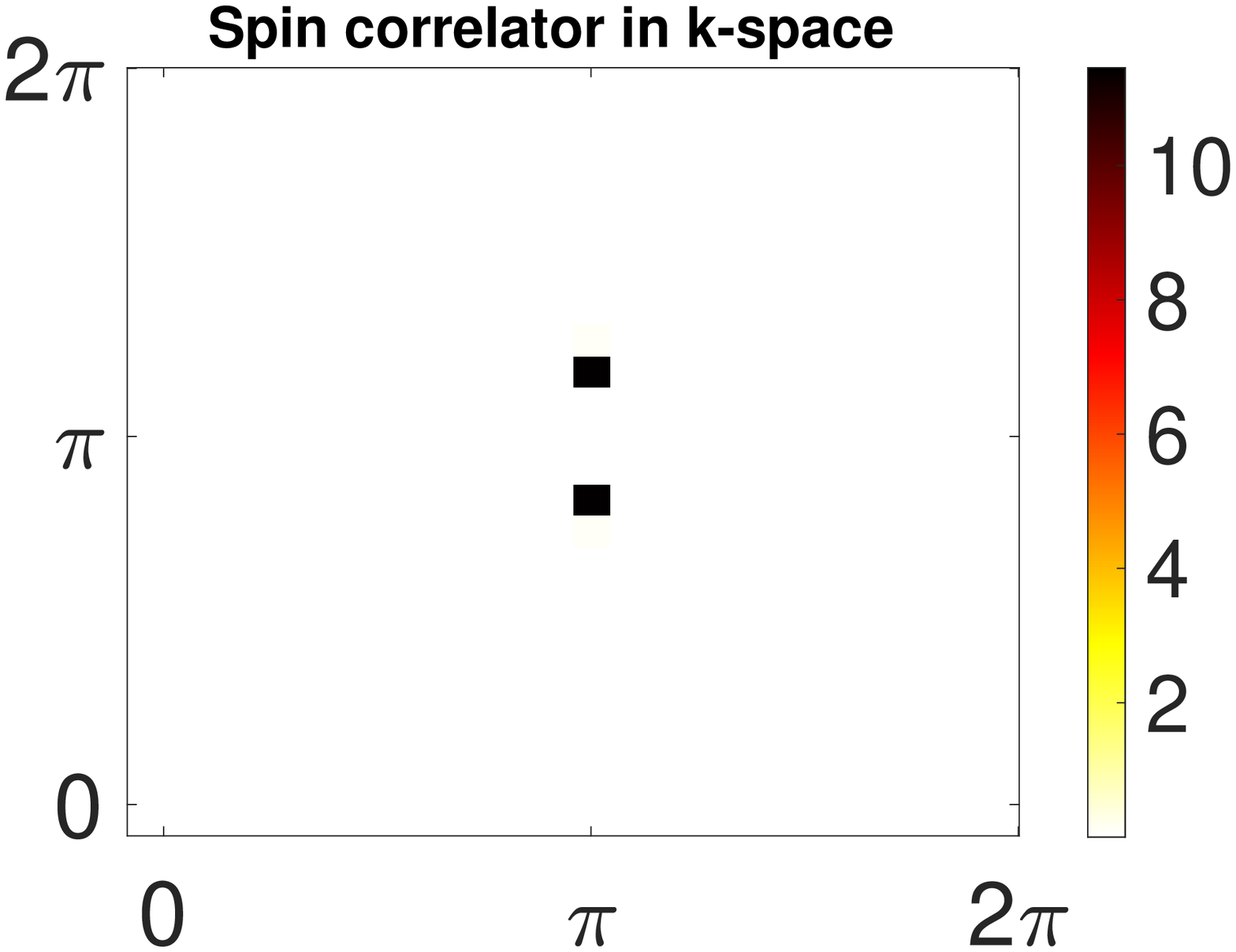}
\label{stripe}
}
\subfigure[~domain]  
{   
 \includegraphics[scale=0.2]{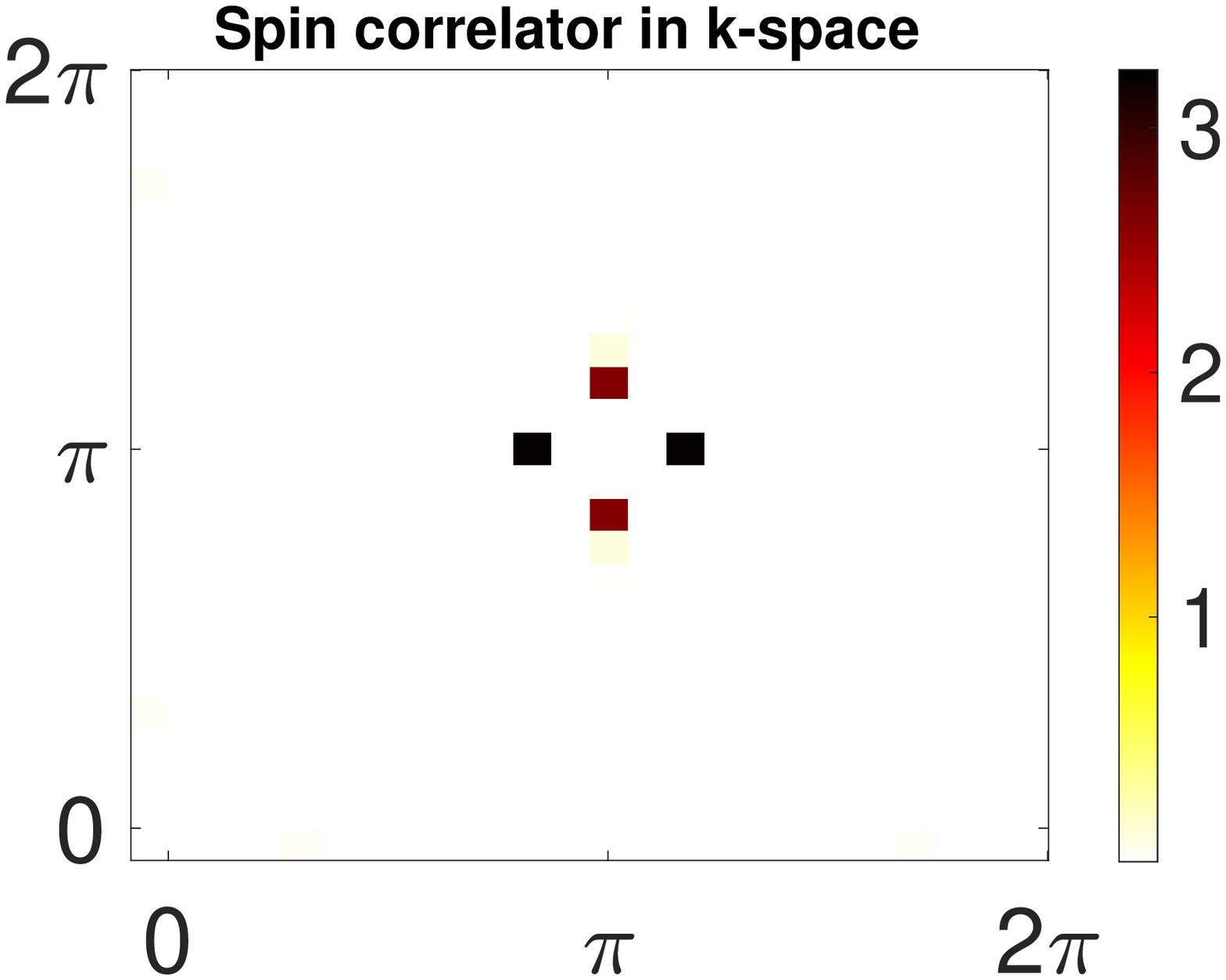}
\label{domain}
}
\subfigure[~checkerboard]  
{   
\includegraphics[scale=0.2]{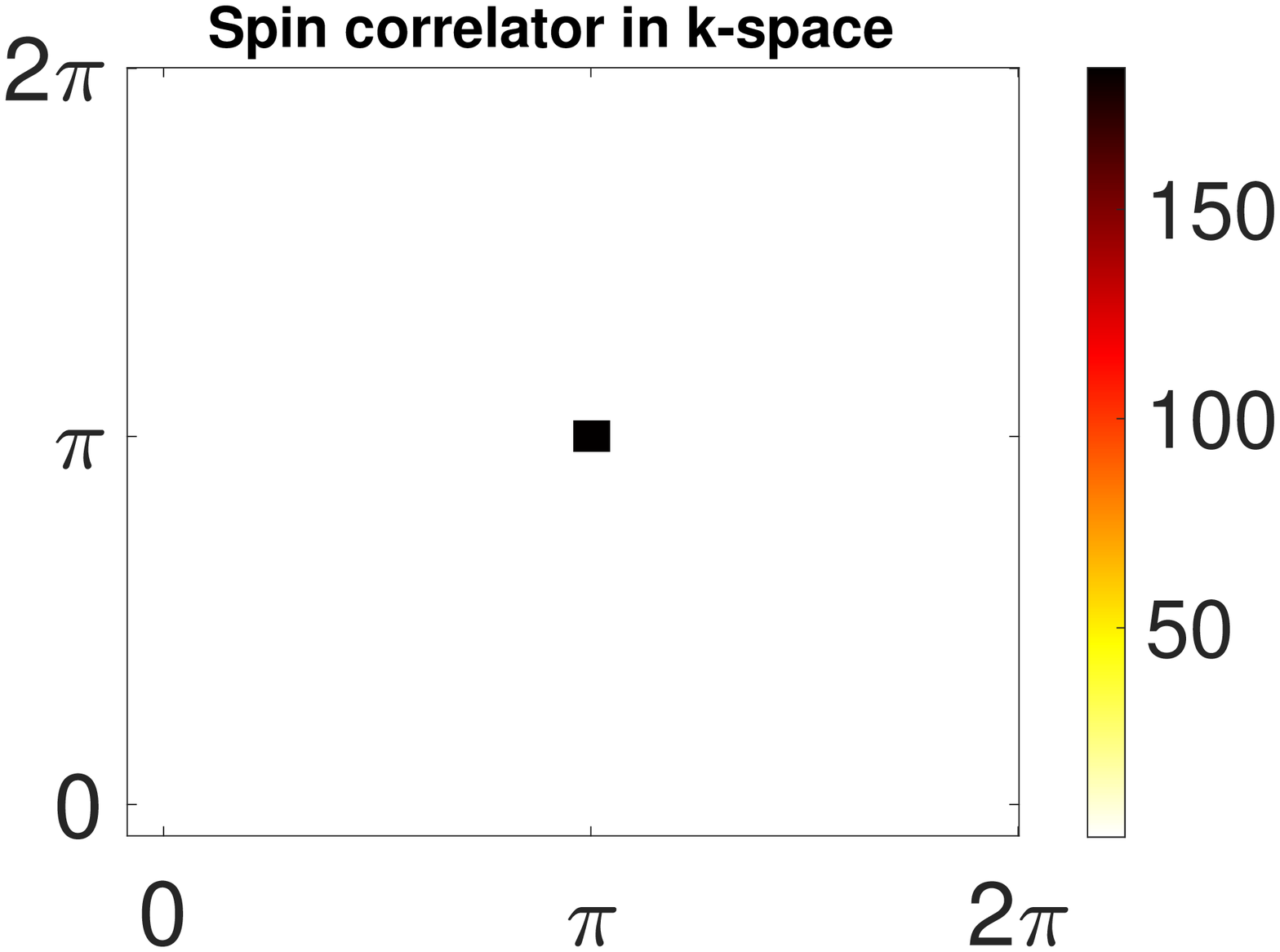}
\label{check}
}
\end{center}
\caption{Varieties of long-range order in momentum space as displayed by the momentum space spin correlator $S(k)$, defined in eq.\ \rf{Sk}, on a $24^2$ lattice at $U=3$ . (a) stripe pattern, $f=0.8$. (b) domain pattern, also at $f=0.8$. (c) checkerboard pattern, $f=1$.}   \label{kspin}
\end{figure}

\begin{figure}[h]
\centerline{\includegraphics[scale=0.6]{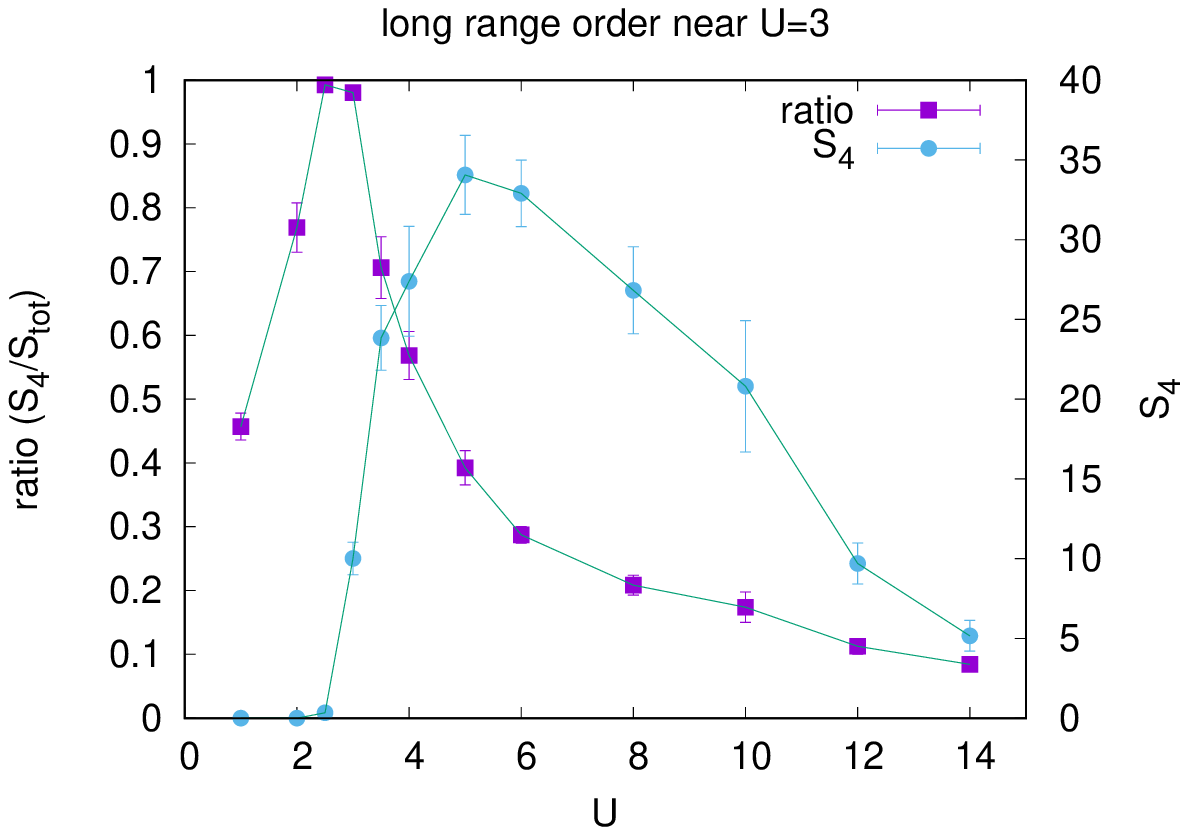}}
\caption{A plot of $S_4/S_{tot}$ (purple squares, left hand $y$-axis), and $S_4$ alone (filled circles, right hand $y$-axis) vs.\ $U$ at
density $f=0.8$.  Here $S_{tot}$ is the sum of all $S(k)$ values, and $S_4$ is the sum of the four largest $k$ values. Long range order
seems most evident near $U=3$, where the sum is almost saturated by the top four values.}
\label{Smax}
\end{figure}

   A more quantitative measure of long range order in magnetization is the momentum space correlator $S(k_x,k_y)$ defined
in \rf{Sk}.  In Fig.\ \ref{kspin} we display this quantity for the stripe, rectangular domain, and checkerboard patterns, corresponding to Figs.\ 
10(b), 10(c), 10(d), 
respectively,
again at $U=3$ on a $24^2$ lattice.   Concentration of $S(k_x,k_y)$ at just a few values (even just one $k_x=k_y=\pi$ for the checkerboard
pattern) is obviously evidence of long range order, and it is of interest to study the $U$-dependence.  So let us choose $f=0.8$, where we see both stripes and rectangular domains, and let $S_4$ be the sum of the largest four values of $S(k_x,k_y)$, with $S_{tot}$ the sum of all
values.  In Fig.\ \ref{Smax} we plot the ratio $S_4/S_{tot}$ (scale on the left-hand y-axis), and $S_4$ alone (scale on the right-hand y-axis).  The long range order at $U=3$ decays away on either side.  In addition, for $U<2$, $S_4$ is hardly distinguishable from zero on the scale of the plot.

\subsection{Pair correlation and the checkerboard}

It is obvious that in an eigenstate of particle number, ${\langle c(k,\up) c(-k,\dn) \rangle = 0}$.  On the other hand, the correlator of
a pair creation operator in the vicinity of point $x$, and a pair destruction operator in the vicinity of point $y$, where $R=|x-y|\gg 1$,
could be non-zero.  Transforming to momentum space, as in eq.\ \rf{pcorr}, it is of special interest to see where in $k$-space this
operator is non-zero, and whether  there is any indication of d-wave symmetry.  In the ground state we find
\bea
          \Delta(k',k) &=& \langle \Phi| c^\dg(k',\up) c^\dg(-k',\dn) c(k,\up) c(-k,\dn)| \Phi \rangle \non \\
                            &=& \sum_{j=2}^M \sum_{i=1}^{j-1} \{ \phi_i(-k',\up)\phi_j(k',\dn) - \phi_i(k',\dn) \phi_j(-k',\up) \} \non \\
                            & & \times \{ \phi_i(k,\up)\phi_j(-k,\dn) - \phi_i(-k,\dn) \phi_j(k,\up) \}  \ .
\eea
Define 
\bea
\o_{ij}(k) &=&  \phi_i(k,\up)\phi_j(-k,\dn) - \phi_i(-k,\dn) \phi_j(k,\up) \non \\
&=&\phi_i(k,\up)\phi^*_j(k,\dn) - \phi^*_i(k,\dn) \phi_j(k,\up) \ .
\eea
Then
\beq
       \Delta(k',k) = \sum_{j=2}^M \sum_{i=1}^{j-1} \o^*_{ij}(k') \o_{ij}(k) \ .
\eeq
The space of all $k,k'$ is four-dimensional, and we choose a two dimensional slice by taking $k'$ to be the wavevector $k$ with $x,y$ 
components interchanged, i.e  $k = (k_x,k_y), k'=(k_y,k_x)$.  If pairing follows a D-wave pattern, then in this two-dimensional slice we would expect something like
\beq
\Delta(k',k) \sim - (\cos(k_x) - \cos(k_y))^2 \ ,
\label{Delta}
\eeq
which is negative everywhere, and most negative at $k_x=0,k_y=\pm \pi$ and $k_x=\pm \pi, k_y=0$.
 
\begin{figure}[htb]
\begin{center}
\subfigure[~f=1, U=1]  
{   
\label{pair1}
 \includegraphics[scale=0.2]{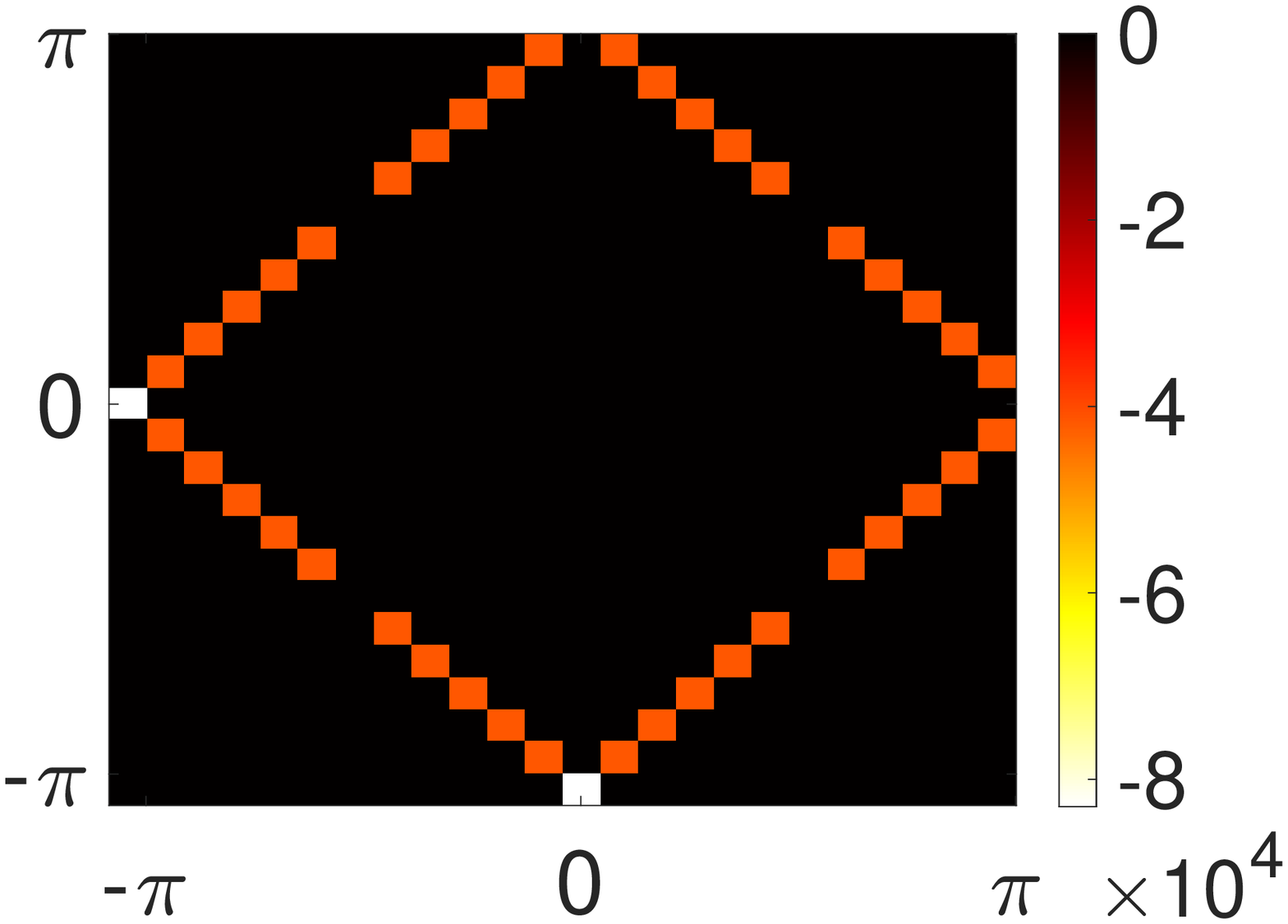}
}
\subfigure[~f=0.95, U=10]  
{   
\label{pair2}
\includegraphics[scale=0.2]{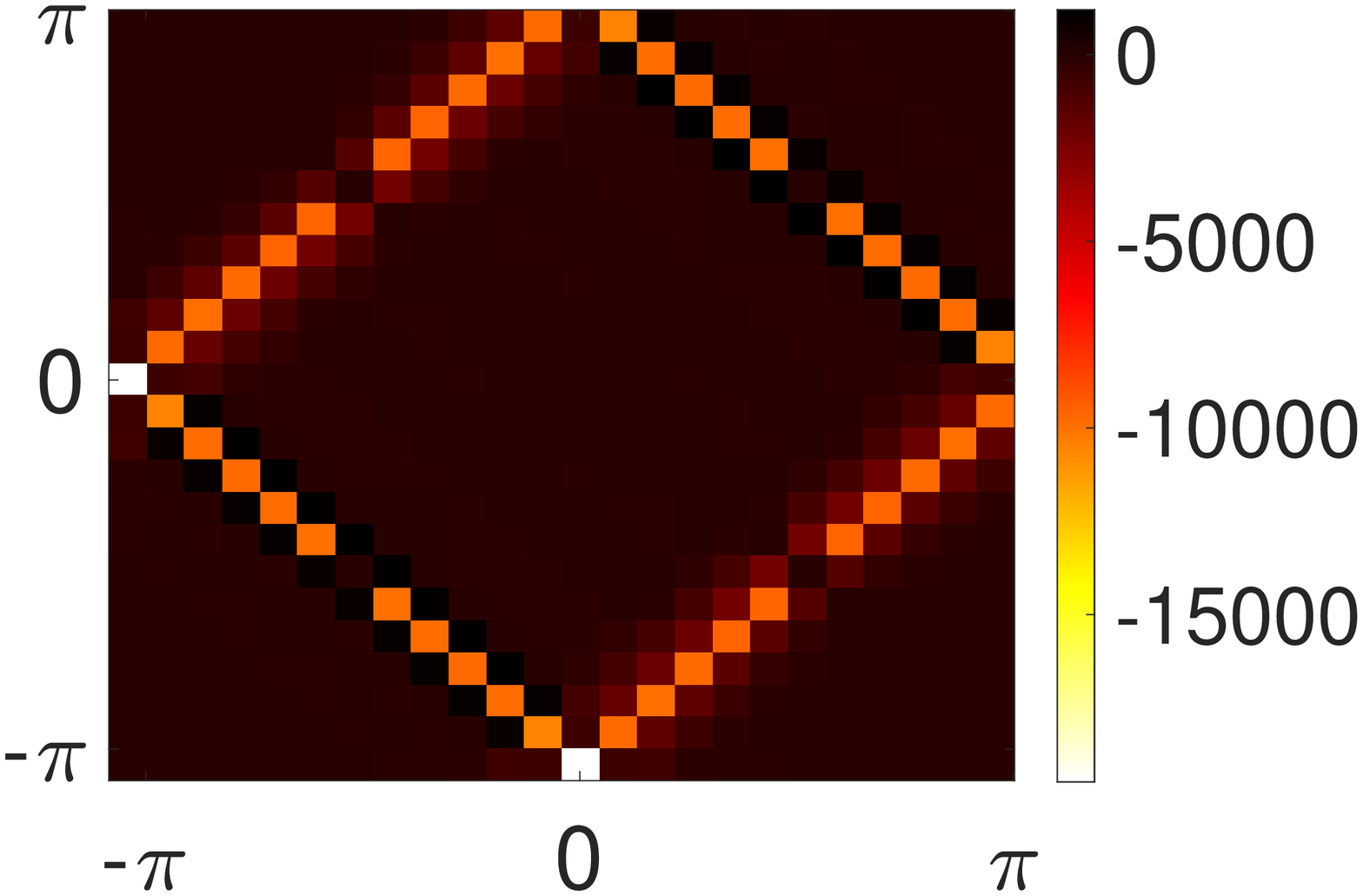}
}
\subfigure[~f=0.8, U=3]
{   
\label{pair3}
\includegraphics[scale=0.2]{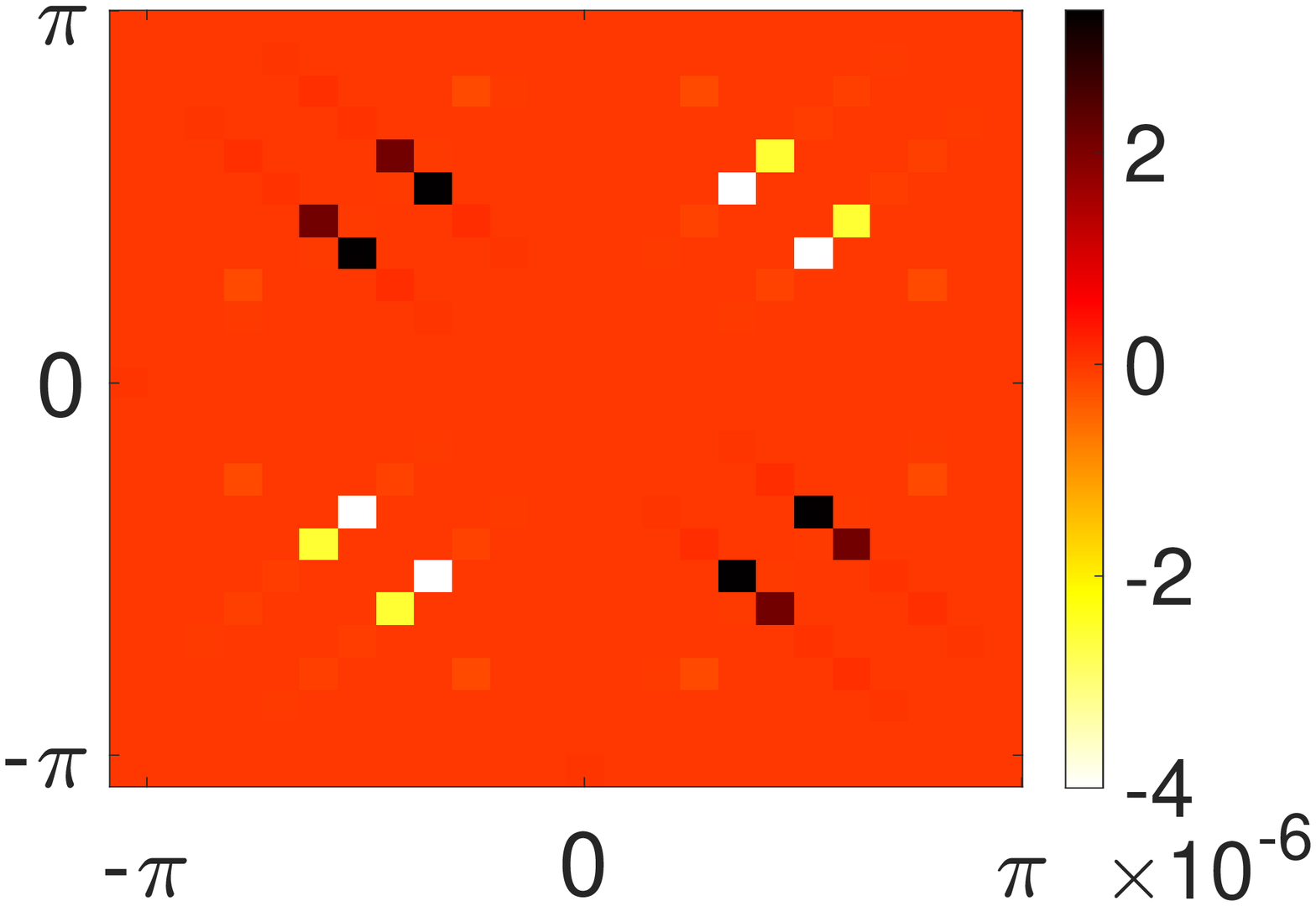}
}
\end{center}
\caption{Display of the pair correlation function $\D(k',k)$ in the $k_x,k_y$ plane, where $k'=(k_x,k_y), k=(k_y,k_x)$.  (a) The pattern shown,
at half-filling and $U=1$, has some of the features of d-wave correlation.  This pattern is virtually unchanged at half-filling across two orders of magnitude, from $U=0.1$ to $10$. (b) Pair correlation at $f=0.95$ and $U=10$. Even slightly away from half-filling, the pattern seen here at $U=10$ is not seen at moderate and small values of $U$. (c) The correlator is negligible (note the scale), compared to the previous subfigures, at $f=0.8, U=3$.  The pattern seen in (a) and (b) is always associated with a checkerboard pattern, and in case (c) there is
instead a stripe pattern.}
\label{pairs}
\end{figure}

   These features are seen, roughly, but {\it only} for checkerboard patterns with the momentum space distribution shown in Fig.\ \ref{pair1},  in the immediate neighborhood of $f=1$. The correlation is also only significant at the edge of the $n(k)$ occupation zone. An example at $f=1,  U/t=1$ is shown in Fig.\ \ref{pair1}, but the very same correlation appears 
at half-filling over a range of two orders of magnitude, from $U/t=0.1$ to $U/t=10$, and the corresponding plots are very similar to Fig.\ \ref{pair1}.\footnote{ We are only interested in the correlator for $k' \ne k$, and have (arbitrarily) set $\Delta(k,k)=0$ in the figures shown.}

The pair correlation at half-filling only starts to disintegrate below $U/t=0.03$.   The situation is quite different away from half-filling, e.g.\ the
checkerboard order and similar pair correlation at $f=0.95$ is only seen at a rather strong repulsion of $U/t=10$ (Fig.\ \ref{pair2}). The picture seen in Figs.\ \ref{pair1} and \ref{pair2} is not exactly what we have in eq.\ \rf{Delta}; for one thing the correlation is sharply concentrated at the boundary of the $n(k)$ occupation zone, and the numerical values do not really agree with \rf{Delta}.
But the fact that the correlator is everywhere negative, and most negative at
$(0,\pm\pi)$ and $(\pm\pi,0)$ is definitely reminiscent of d-wave correlation.  We emphasize, however, that the checkerboard pattern and this type of pair correlation are only found in the immediate neighborhood of half-filling, and even then (e.g.\ at $f=0.95$) may only be seen at comparatively large values of $U/t$, as in Fig.\ \ref{pair2}.  When the checkerboard pattern is absent the correlation function $\D(k,k')$ is still generally concentrated at the edge of the $n(k)$ distribution, but may be everywhere positive, or a mixture of positive and negative values.  Or the pair correlation may be negligible, as seen in Fig.\ \ref{pair3} (note the scale) at $f=0.8, U=3$, 
where there is a stripe order.\footnote{We note that a weak-coupling analysis of the 2D Hubbard model in ref.\ \cite{Ragu} also found d-wave pairing strongest near half-filling.}
 
\subsection{Localization and particle-hole symmetry}

\begin{figure}[t!]
\centerline{\includegraphics[scale=0.25]{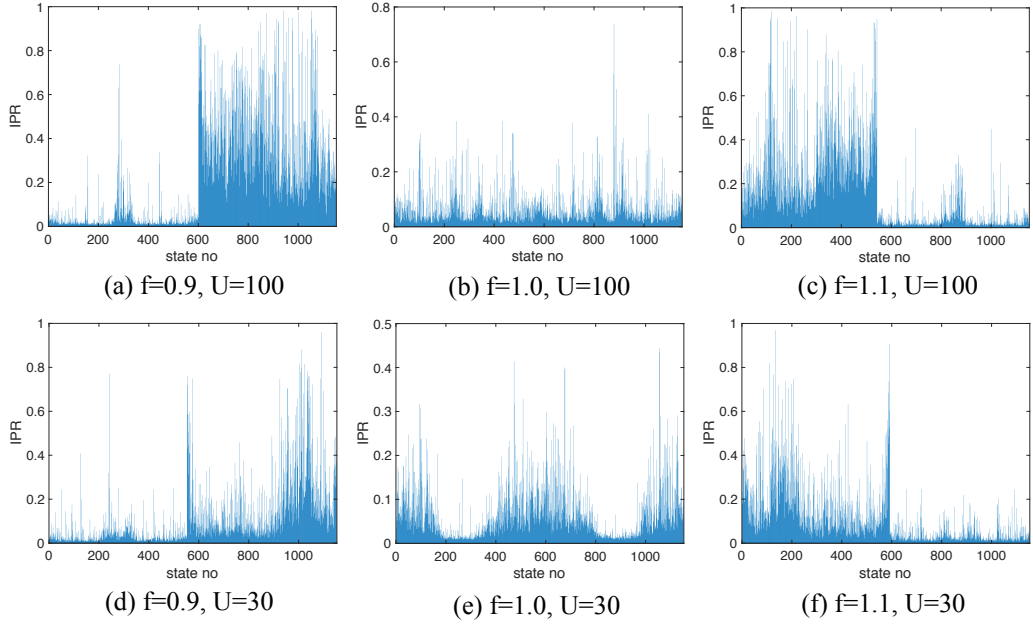}}
\caption{Localization of particle and hole states at very strong repulsion $U=30,100$ as quantified by their IPR values.  The $x$-axis is the one-particle state number, i.e.\ the $i$ of $\phi_i(x,s)$.
Note that it is hole states which are localized for $f<1$, and particle states which are localized for $f>1$, with a symmetric distribution
at $f=1$.  Subfigures (a,b,c): $U=100$ and $f=0.9,1.0,1.1$ respectively. Subfigures (d,e,f): $U=30$ and $f=0.9,1.0,1.1$ respectively.}
\label{IPR100} 
\end{figure}

   We return, finally, to the question which motivated this work. The stripe, domain and checkerboard patterns do not imply that one electron wave functions are localized, and in fact the opposite is
true, as we have already noted.  The Hartree-Fock state may break translation invariance, as is obvious from the geometric patterns,
but according to our investigation of the IPR values of the constituent one-particle wave functions in $|\Phi\rangle$, these wave functions
spread over most of the lattice. The situation is different at strong $U/t$, as
can be seen in Fig.\ \ref{IPR100}, at  densities $f=0.9,1.0,1.1$ and $U/t=30,100$.  In this figure we display the IPR values of all $2L^2$ particle states ($L=24$), both filled and unfilled.  An interesting feature is that it is the hole (unfilled) states which are localized at $f<1$, and the electron (filled) states which are localized at $f>1$, with about an equal degree of localization among hole and particle states at $f=1$.  The dramatic shift from localized holes at $f<1$ to localized electrons at $f>1$, no doubt a consequence of particle-hole symmetry, is most obvious at an extremely strong coupling $U/t=100$, but it also quite evident at, e.g., $U/t=30$.  As $U/t$ is further reduced to moderate values, the asymmetry in particle/hole localization on either side of $f=1$, and the localization itself, gradually disappears.

\begin{figure}[htb]
\begin{center}
\subfigure[~$\d=10^{-3}$]  
{   
\label{d3}
 \includegraphics[scale=0.25]{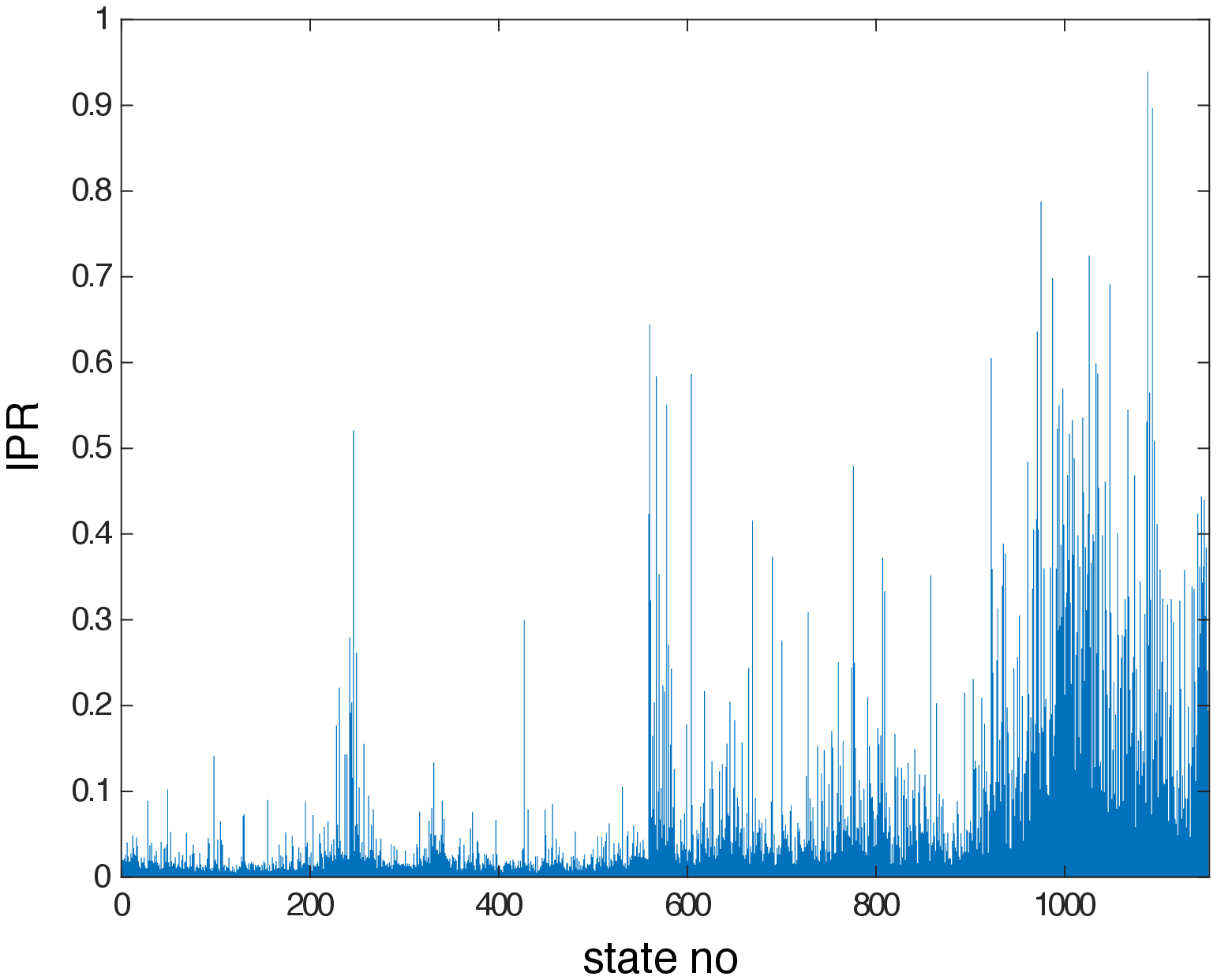}
}
\subfigure[~$\d=10^{-4}$]  
{   
\label{d4}
\includegraphics[scale=0.25]{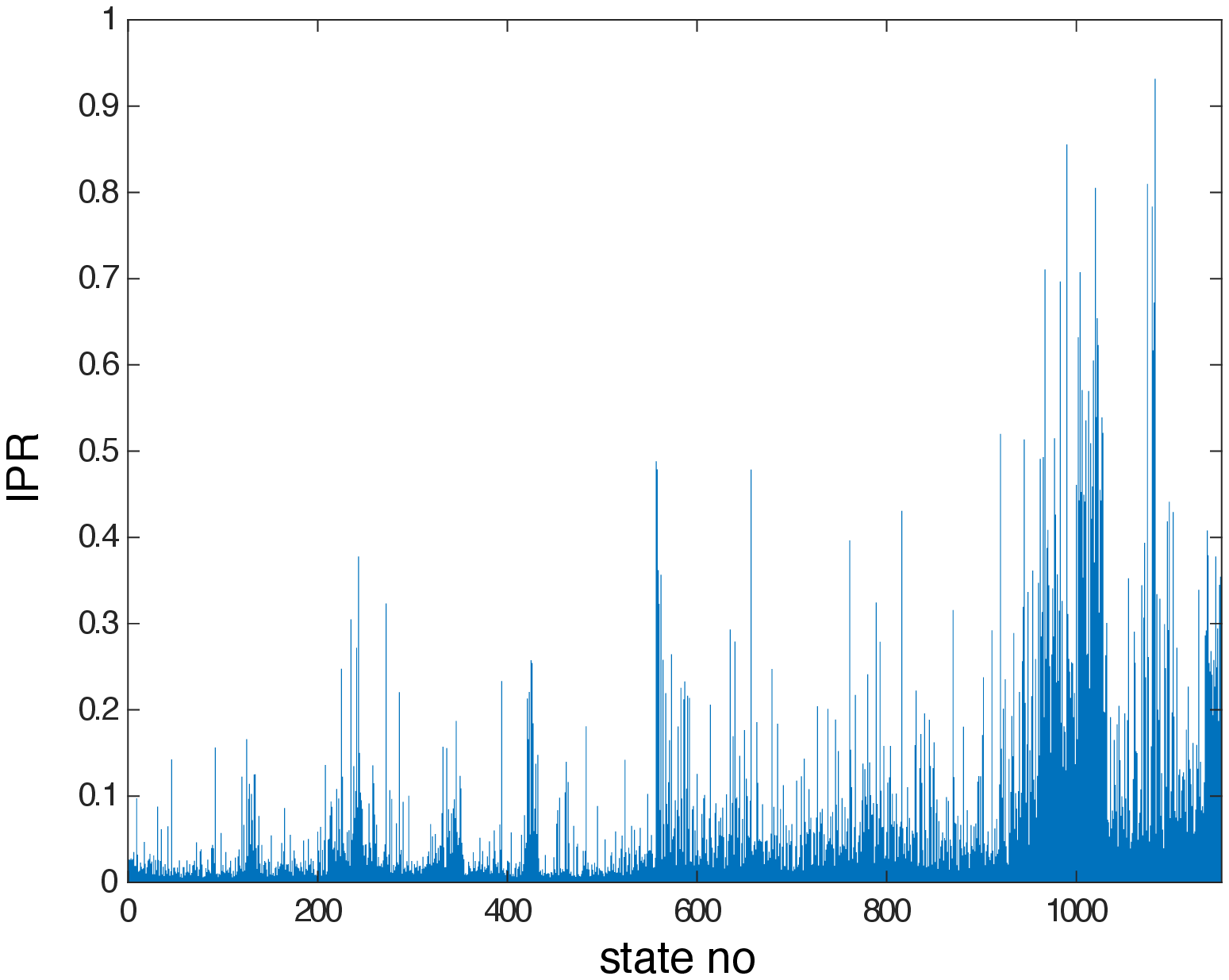}
}
\subfigure[~$\d=10^{-5}$]
{   
\label{d5}
\includegraphics[scale=0.25]{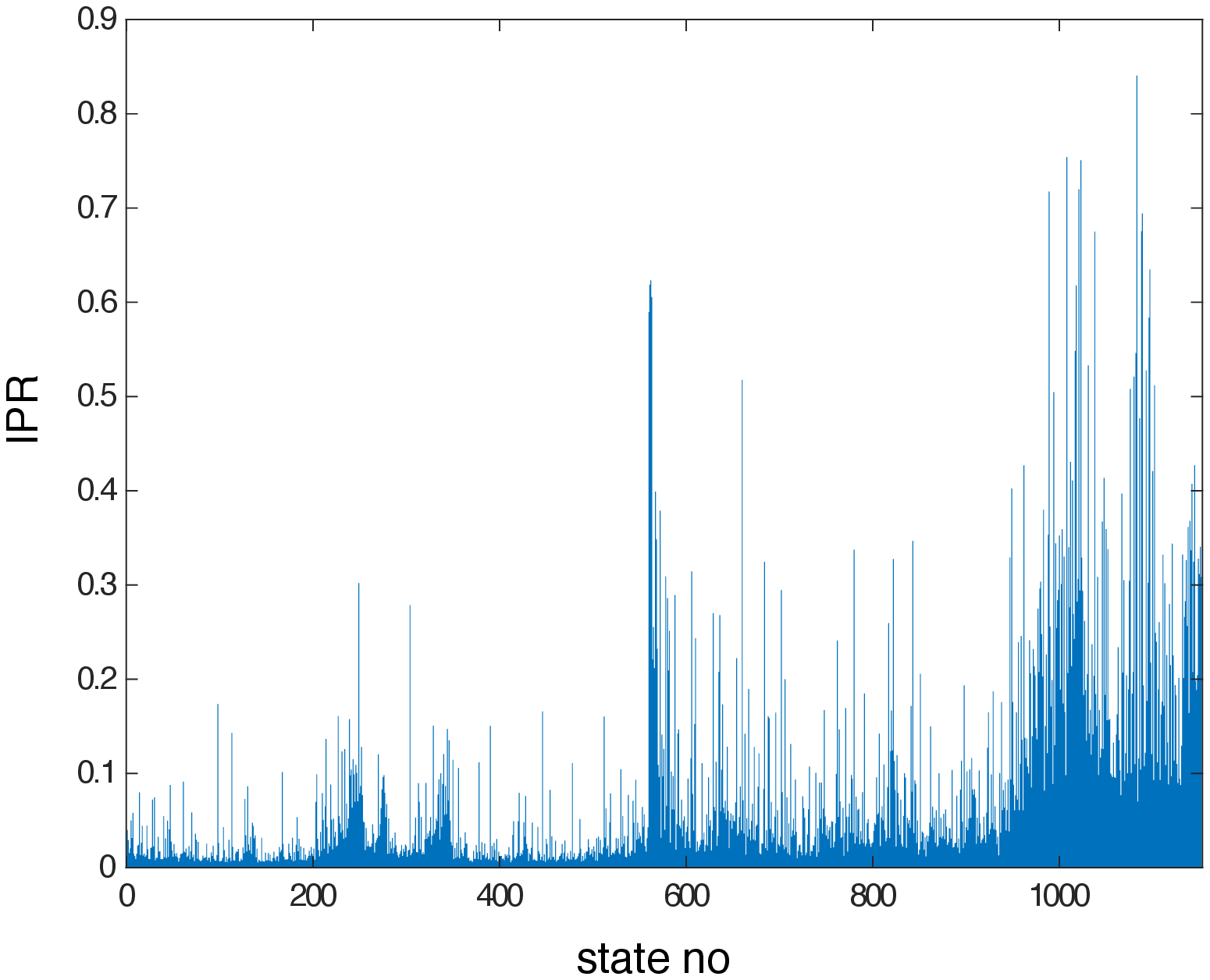}
}
\end{center}
\caption{A comparison of IPR values at $U=30, f=0.9$ (same as Fig.\ 15(d)), but with different convergence
criteria parametrized by $\d$.  Each figure is taken from a different self-consistent solution.  While there is some variation
in the IPR values shown in these figures, it is consistent with the variation at fixed $\d=10^{-3}$ among different
self-consistent solutions.}
\label{deltas}
\end{figure}

   It is a general feature of the iterative procedure that the number of iterations required for convergence increases with $U/t$, and it is
of interest to repeat the calculation with more stringent convergence criteria.  
The computation of IPR values shown in Fig.\ \ref{IPR100}(d), with
convergence parameter $\d=10^{-3}$ was repeated at  $\d=10^{-4}$ and $\d=10^{-5}$.  At  $U/t=30, f=0.9$, the number of iterations required for convergence increases very significantly, with 2030 iterations required at $\d=10^{-3}$, 10810 iterations at $\d=10^{-4}$, and
40510 at $\d=10^{-5}$.  Nevertheless, the qualitative picture of localization is unchanged by making the convergence criterion
more stringent.   In Fig.\ \ref{deltas} we compare IPR values at $U=30, f=0.9$ for different values of the convergence
parameter $\d$, where each subfigure is derived from a different self-consistent solution of the Hartree-Fock equations.  Here it should be understood that there is always some variation in the IPR values of one-particle eigenstates $\phi_i$ from one self-consistent solution to the next, and the variation seen by decreasing $\d$ is not much different from the variation among different self-consistent 
solutions at fixed $\d$. 
 
Anderson localization is a well known phenomenon for particle propagation in a random potential, and we feel it is significant that 
here, by contrast, there is localization in a translationally invariant system, where the degree of localization
among the different states depends on the random initialization.  This seems to answer in the affirmative the question raised
in \cite{Nandi}, which asks whether this phenomenon is possible.

\section{Conclusions}

   The Hartree-Fock approach yields a multiplicity of self-consistent solutions of surprising complexity; more complexity than one might expect given that any Hartree-Fock state is just a single Slater determinant, where entanglement is limited to what is required by Fermi statistics.  We see emergent spin patterns in the form of stripes, checkerboards, and rectangular domains, band gaps correlated with antiferromagnetic order, localization of particles or holes for strong repulsion at $f>1$, $f<1$ respectively, and pair correlation functions near half-filling which have properties similar to d-wave distributions.   Perhaps most important is the multiplicity itself; i.e.\ the enormous number self-consistent solutions of the Hartree-Fock equations, derived from an iterative procedure with only slightly different random initializations, that are very nearly degenerate in energy.  

    One can take different views of this near degeneracy.  One view is that the true ground state, so far as it can be determined in the Hartree-Fock approach, is the state with the lowest energy.  This is in a landscape of states which have very nearly the same energy, but are distinguished from one another by other observables, such as local magnetization and energy gap, which can differ widely among different solutions. If it were true that the low temperature behavior is dominated by a unique lowest energy ground state, then in our opinion this
would rule out translation non-invariance in the expectation value of any observable.   The other possible view, if we take the multiplicity of near degenerate states seriously, is that as the system is cooled from finite temperature it finds itself in one of those near-degenerate states,  
rarely if ever falling into the true quantum ground state.  If that is so, then the 2D Hubbard model may have other features analogous to spin glasses, e.g.\ some degree of non-ergodicity, which deserve further study.

    We have also found localization of one-particle hole and electron states just below and just above half-filling, at strong repulsion, with IPR values which again depend on the particular solution obtained from the particular initialization.  Unlike in ordinary Anderson localization there is no random potential intrinsic to the many body Hamiltonian, but a random effective potential, as seen by a single particle, may arise from the random initialization.  There has been speculation \cite{Nandi}  whether small random differences in intialization might drive a system to become localized, and our results would seem to support that possibility.

    The Hartree-Fock approach to the Hubbard model must obviously be viewed with caution; as in any mean-field theory it simply ignores the
correlation of the average field acting on one particle with the position/spin of that particle, and this neglect is well known to lead to errors, particularly in the neighborhood of a transition.  In its defense, we have seen that apart from very strong repulsion, the one-electron wave functions are unlocalized.  That means that a single electron is effectively interacting with all other electrons in the system, and not just with a handful of nearest neighbors. The fact that one degree of freedom is interacting with many is the usual mean-field (and in particular Hartree-Fock) justification for replacement of the ``many'' by their average.   On the other hand, since entanglement in a Hartree-Fock state is limited to what is required by Fermi statistics, any phenomenon which depends on entanglement beyond that requirement will simply be invisible
in the framework of the Hartree-Fock approximation.  Among the important observables which may depend crucially on entanglement we
would include electron pairing.   

    All of the work presented in this article concerns the zero temperature Hartree-Fock approximation to the 2D Hubbard model. We leave an investigation of localization by more sophisticated methods, perhaps including the effects of finite temperature, to later investigation.
 
\bigskip

\ni {\bf Acknowledgements} \\
This work is supported by the U.S.\ Department of Energy under Grant No.\ DE-SC0013682.

\bibliography{hub}
\end{document}